\documentclass[final,comsoc]{IEEEtran}

\usepackage{amsmath,amssymb}
\usepackage{cite}
\usepackage{graphicx}
\usepackage{paralist}
\usepackage{tikz} \usetikzlibrary{positioning,shapes.geometric,arrows,patterns}

\newtheorem{theorem}{Theorem}
\newtheorem{proposition}{Proposition}
\newtheorem{lemma}{Lemma}
\newtheorem{definition}{Definition}

\graphicspath{{./fig/}}

\DeclareMathOperator{\subjectto}{subject~to}

\DeclareMathOperator*{\minimize}{minimize}

   \def\bx{{\boldsymbol{x}}}
\def\be{{\boldsymbol{e}}}  \def\bs{{\boldsymbol{s}}} \def\by{{\boldsymbol{y}}}

 \def\bK{{\boldsymbol{K}}}  \def\bX{{\boldsymbol{X}}}
  \def\bS{{\boldsymbol{S}}} \def\bY{{\boldsymbol{Y}}}
   \def\bZ{{\boldsymbol{Z}}}

   \def\bxx{{\underline{\boldsymbol{x}}}}
  \def\bss{{\underline{\boldsymbol{s}}}}

   \def\bXX{{\underline{\boldsymbol{X}}}}
  \def\bSS{{\underline{\boldsymbol{S}}}}

\def\prob{\mathsf{P}}
\def\Exp{\mathsf{E}}
\def\Bcal{\mathcal{B}}
\def\Ecal{\mathcal{E}}
\newcommand{\grp}{G}

\newcommand{\icol}[1]{
  \left(\begin{smallmatrix}#1\end{smallmatrix}\right)%
}

\begin{document}
%
\title{Capacity of Gaussian Many-Access Channels}
%
%
%
\author{
  \IEEEauthorblockN{Xu Chen, Tsung-Yi Chen, and Dongning Guo}
  \thanks{X. Chen was with the Department of Electrical Engineering and Computer Science, Northwestern University, Evanston, IL. He is now with Apple Inc., Cupertino, CA.

 T.-Y. Chen was with the Department of Electrical Engineering and Computer Science, Northwestern University, Evanston, IL. He is now with SpiderCloud Wireless Inc., San Jose, CA.

D. Guo is with the Department of Electrical Engineering and Computer Science, Northwestern University, Evanston, IL.  dGuo@Northwestern.edu.

  This work was presented in part at the 2013 IEEE Information Theory Workshop, Sevilla, Spain~\cite{chen2013gaussian} and the 2014 IEEE International Symposium on Information Theory, Honolulu, HI~\cite{chen2014manyaccess}.

 This work was supported in part by the National Science Foundation under Grant Nos. ECCS-1231828 and CCF-1423040.}
}

\maketitle

\begin{abstract}
Classical multiuser information theory studies the fundamental limits of models with a fixed (often small) number of users as the coding blocklength goes to infinity. This work proposes a new paradigm, referred to as {\em many-user information theory}, where the number of users is allowed to grow with the blocklength. This paradigm is motivated by emerging systems with a massive number of users in an area, such as the Internet of Things.  The focus of the current paper is the {\em many-access} channel model, which consists of a single receiver and many transmitters, whose number increases unboundedly with the blocklength. Moreover, an unknown subset of transmitters may transmit in a given block and need to be identified as well as decoded by the receiver. A new notion of capacity is introduced and characterized for the Gaussian many-access channel with random user activities. The capacity can be achieved by first detecting the set of active users and then decoding their messages.  The minimum cost of identifying the active users is also quantified.
\end{abstract}

\section{Introduction}
\label{sec:intro}

Classical information theory characterizes the fundamental limits of communication systems by studying the asymptotic regime of infinite coding blocklength. The prevailing models in multiuser information theory assume a fixed (usually small) number of users, where fundamental limits as the coding blocklength goes to infinity are studied. Even in the large-system analysis of multiuser systems~\cite{gupta2000capacity,verdu1999spectral,guo2005randomly}, the blocklength is sent to infinity before the number of users is sent to infinity.\footnote{The same can be said of the many-user broadcast coding strategy for the point-to-point channel proposed in~\cite{shamai1997broadcast}, and the CEO problem~\cite{berger1996ceo}.} In some sensor networks and emerging Internet of Things, a massive and ever-increasing number of wireless devices with sporadic traffic may need to share the spectrum in a given area. This motivates us to rethink the assumption of fixed population of fully buffered users. Here we propose a new {\em many-user paradigm}, where the number of users is allowed to increase without bound with the blocklength.\footnote{The only existing model of this nature is found in~\cite{chang1979coding}, in which the authors sought for uniquely-decodable codes for a noiseless binary adder channel where the number of users increases with the blocklength.}

In this paper, we introduce the many-access channel~(MnAC) to model systems consisting of a single receiver and many transmitters, the number of which is comparable to or even larger than the blocklength~\cite{chen2013gaussian,chen2014manyaccess}. We study the asymptotic regime where the number of transmitting devices ($k$) increases as the blocklength ($n$) tends to infinity. The model also accommodates random access, namely, it allows each transmitter to be active with certain probability in each block.
We assume synchronous transmission in the model.\footnote{A recent follow-up work~\cite{shahi2016capacity} has studied the capacity of strong asynchronous MnACs.}

In general, the classical theory does not apply to systems where the number of users is comparable or larger than the blocklength, such as in a machine-to-machine communication system with many thousands of devices in a given cell. One key reason is that, for many functions of two variables $f$, $\lim_{k \to \infty} \lim_{n \to \infty} f(k,n) \neq \lim_{n \to \infty} f(k_n, n)$, i.e., letting $k \to \infty$ after $n \to \infty$ may yield a different result than letting $n$ and $k = k_n$ (as a function of $n$) simultaneously tend to infinity. Moreover, the traditional notion of rate in bits per channel use is ill-suited for the task in the many-user regime as noted (for the Gaussian multiaccess channel) in~\cite[pp.~546--547]{cover2006elements} by Cover and Thomas, ``when the total number of senders is very large, so that there is a lot of interference, we can still send a total amount of information that is arbitrary large even though the rate per individual sender goes to 0.''

Capacity of the conventional multiaccess channel is well understood~\cite{ahlswede1971multiway,liao1972coding,gallager1985perspective}. The achievable error exponent and capacity region of a random multiaccess channel were derived in~\cite{plotnik1991decoding}. Packet-based random multiaccess communication systems with collision detection have also been studied from the perspective of information theory in~\cite{ wang2012error, luo2012new}. The capacity of the conventional multiaccess channel can be established using the fact that joint typicality is satisfied with probability 1 as the blocklength grows to infinity. This argument, however, does not directly apply to models where the number of users also goes to infinity. Specifically, joint typicality requires the simultaneous convergence of the empirical joint entropy of every subset of the input and output random variables to the corresponding joint entropy. Even though convergence holds for every subset due to the law of large numbers, the asymptotic equipartition property is not guaranteed because the number of those subsets increases exponentially with the number of users~\cite{chen2014manybroadcast}. Resorting to strong typicality does not resolve this because the empirical distribution over an increasing alphabet (due to increasing number of users) does not converge.

In general, the received signal of the Gaussian MnAC is a noisy superposition of the codewords chosen by the active users from their respective codebooks. The detection problem boils down to identifying codewords based on their superposition. It is closely related to sparse recovery, also known as compressed sensing, which has been studied in a large body of works~\cite{donoho2006compressed,candes2006near,candes2005decoding,wainwright2009information,wainwright2009sharp,fletcher2009necessary,wang2010information,akcakaya2010shannon,aeron2010information,rahnama2011nearly}.
Information-theoretic limits of exact support recovery was considered in~\cite{wainwright2009information}, and stronger necessary and sufficient conditions have been derived subsequently~\cite{fletcher2009necessary,wang2010information,rahnama2011nearly}.
Using existing results in the sparse recovery literature, it can be shown that the message length (in bits) that can be transmitted reliably by each user through the MnAC should be in the order of $\Theta(  n (\log k_n) / k_n )$.

In this paper, we provide a sharp characterization of the capacity of Gaussian many-access channels as well as the user identification cost.
As an achievable scheme, each user's transmission consists of a signature that identifies the user, followed by a message-bearing codeword. The decoder first identifies the set of active users based on the superposition of their unique signatures. (This is in fact a compressed sensing problem~\cite{zhang2013neighbor,zhang2014virtual}.) It then decodes the messages from the identified active users.
The length of the signature matches the capacity penalty due to user activity uncertainty. The proof techniques find their roots in Gallager's error exponent analysis~\cite{gallager1968information}. Also studied is a more general setup where groups of users have heterogeneous channel gains and activity patterns. Again, separate identification and decoding is shown to achieve the capacity region.
While the exact capacity of the MnAC with given large finite user population and blocklength remains a hard open problem, this paper offers a new asymptotic theory that has a better explanatory power for random massive access than the classical theory.

Unless otherwise noted, we use the following notational conventions: $x$ denotes a scalar, $\bx$ denotes a column vector, and $\bxx$ denotes a matrix. The corresponding uppercase letters $X$, $\bX$, and $\bXX$ denote the corresponding random scalar, random vector and random matrix, respectively. Given a set $A$, let $\bx_A = (x_i)_{i \in A}$ denote the subset of variables of $\bx$ whose indices are in $A$ and let $\bxx_A = (\bx_i)_{i \in A}$ be the matrix formed by columns of $\bxx$ whose indices are in $A$. Let $x_n \leq_{n} y_n$ denote $\limsup_{n \to \infty} (x_n - y_n) \leq 0$.
That is, $x_n$ is essentially asymptotically dominated by $y_n$. All logarithms are natural. The binary entropy function is denoted as $H_2(p) = -p \log p -(1-p) \log (1-p)$.

The rest of this paper is organized as follows.  Section~\ref{sec:systemmodel} presents the system model and main results.
Section~\ref{sec:converse} proves the converse part of the MnAC capacity result. Section~\ref{sec:proof_capacity_identify} quantifies the user identification cost. Section~\ref{sec:achievability} proves the achievability part of the MnAC capacity result. Section~\ref{sec:succ_decode} discusses successive decoding techniques for MnAC.  Section~\ref{sec:hetergain} analyzes the capacity of MnAC with heterogeneous user groups.  Concluding remarks are given in Section~\ref{sec:conclude}.

\section{System Model and Main Results}
\label{sec:systemmodel}

Let $n$ denote the number of channel uses, i.e., the blocklength.  Let the total number of users be tied to the blocklength and be denoted as $\ell_n$, which is a function of $n$.  The received symbols in a block form a column vector of length $n$:
\begin{align}\label{eq:systemmodel}
\bY = \sum_{k =1 }^{\ell_n} \bS_k (w_k) + \bZ
\end{align}
where $w_k$ is the message of user $k$, $\bS_k (w_k) \in \mathbb{R}^{n }$ is the corresponding $n$-symbol codeword, and $\bZ $ is a Gaussian noise vector with independent standard Gaussian entries. Suppose each user accesses the channel independently with identical probability $\alpha_n$ during any given block. If user $k$ is inactive, it is thought of as transmitting the all-zero codeword $\bs_k(0) = \mathbf{0}$.

\begin{definition}\label{def:MNCode}
Let $\mathcal{S}_k$ and $\mathcal{Y}$ denote the input alphabet of user $k$ and the output alphabet of the MnAC, respectively. An $(M,n)$ symmetric code with power constraint $P$ for the MnAC channel $(\mathcal{S}_1 \times \mathcal{S}_2 \times \cdots \times \mathcal{S}_{\ell_n}, p_{Y|S_1, \cdots, S_{\ell_n}},\mathcal{Y})$ consists of the following mappings:
\begin{enumerate}
  \item The encoding functions $\Ecal_k: \{ 0,1,\dots,M \} \rightarrow \mathcal{S}_k^n$ for every user $k \in \{ 1, \cdots, \ell_n \}$, which maps any message $w$ to the codeword $\bs_k (w) =[s_{k1}(w), \cdots, s_{kn} (w) ]^T$. In particular, $\bs_k(0) = \mathbf{0}$, for every $k$. Every codeword $\bs_k (w)$ satisfies the power constraint:
      \begin{align} \label{eq:powerconst}
       \frac{1}{n} \sum_{i=1}^n s_{ki}^2 (w) \leq P .
    \end{align}

  \item Decoding function $\mathcal{D}: \mathcal{Y}^n \rightarrow \{0, 1,\dots,M \}^{\ell_n}$, which is a deterministic rule assigning a decision on the messages to each possible received vector.
\end{enumerate}

The average error probability of the $(M,n)$ code is:
\begin{align}\label{eq:Perror}
  \prob_e^{(n)} = \prob \left\{ \mathcal{D}(\bY) \neq (W_1, \dots, W_{\ell_n}) \right\},
\end{align}
where the messages $W_1, \cdots, W_{\ell_n} $ are independent, and for $k \in \{1, \cdots, \ell_n\}$, the message's distribution is
\begin{align}
\prob \left\{ W_k= w \right\} =
\begin{cases}
1-\alpha_n, \quad & w=0, \\
\frac{\alpha_n}M, \quad & w \in\{1,\dots,M\}.
\end{cases}
\end{align}
\end{definition}

The code is said to be {\em symmetric} because the message length is the same for all users. (We extend to an asymmetric case in Section~\ref{sec:hetergain}.) The preceding model reduces to the conventional $\ell$-user multiaccess channel in the special case where $\ell_n = \ell$ is fixed and $\alpha_n =1$ as the blocklength $n$ varies.

\subsection{The Message-Length Capacity}
\label{sec:message_capacity}

\begin{definition}[Asymptotically achievable message length]\label{def:msglen}
We say a positive nondecreasing sequence of message lengths $\left\{ v(n) \right\}_{n=1}^{\infty}$, or simply, $v(\cdot)$, is asymptotically achievable for the MnAC if there exists a sequence of $(\lceil \exp(v(n)) \rceil,n)$ codes according to Definition~\ref{def:MNCode} such that the average error probability $\prob_e^{(n)}$ given by~\eqref{eq:Perror} vanishes as $n \to \infty$.
\end{definition}

It should be clear that by asymptotically achievable message length we really mean a function of the blocklength. The base of $\exp(\cdot)$ should be consistent with the unit of the message length. If the base of $\exp(\cdot)$ is 2 (resp. $e$), then the message length is measured in bits (resp. nats).

\begin{definition}[Symmetric message-length capacity]\label{def:symmetriccapacity}
For the MnAC channel described by~\eqref{eq:systemmodel}, a positive nondecreasing function $B(n)$ of the blocklength $n$ is said to be a symmetric message-length capacity of the MnAC channel if, for any $0 <\epsilon <1$, $ (1 - \epsilon) B(n) $ is an asymptotically achievable message length according to Definition~\ref{def:msglen}, whereas $ (1+\epsilon) B(n) $ is not asymptotically achievable.
\end{definition}

For the special case of a (conventional) multiaccess channel, the symmetric capacity $B(n)$ in Definition~\ref{def:symmetriccapacity} is asymptotically linear in $n$, so that $\lim_{n \to \infty} B(n) / n$ is equal to the symmetric capacity of the multiaccess channel (in, e.g., bits per channel use). From this point on, by ``capacity'' we mean the message-length capacity in contrast to the conventional capacity.

Definition~\ref{def:symmetriccapacity} is only concerned with the asymptotics of the message length.  If $B(n)$ is a capacity, then so is $B(n)+o(B(n))$.  Hence the capacity expression is not unique.
In general, the message-length capacity $B(n)$ need not grow linearly with the blocklength.

Let $\bSS_k = [\bS_k(1) , \cdots, \bS_k(M)]$ denote the matrix consisting of all but the first all-zero codeword of user $k$. Let $\bSS = [\bSS_1, \cdots, \bSS_{\ell_n}] \in \mathbb{R}^{n \times( M \ell_n)}$ denote the concatenation of the codebooks of all users. For ease of analysis, we often use the following equivalent model for the Gaussian MnAC~\eqref{eq:systemmodel}:
\begin{align}\label{eq:systemmodel2}
\bY = \bSS \bX + \bZ,
\end{align}
where 
$\bZ $ is defined as in~\eqref{eq:systemmodel} and $\bX \in \mathbb{R}^{M \ell_n}$ is a vector indicating the codewords transmitted by the users. Specifically, $\bX = [\bX_1^T, \bX_2^T, \cdots , \bX_{\ell_n}^T]^T$, where $\bX_k \in \mathbb{R}^M$ indicates the codeword transmitted by user $k$, $k=1,\cdots, \ell_n$, i.e.,
\begin{align}
  \bX_k =
  \begin{cases}
    \mathbf{0}\;\; & \text{ with probability } 1- \alpha_n \\
    \be_m & \text{ with probability } \frac{\alpha_n}{M}, \enskip m = 1, \dots, M
  \end{cases}
\end{align}
where $\be_m$ is the binary column $M$-vector with a single 1 at the $m$-th entry. Let
\begin{align}
\begin{split}
  \mathcal{X}^{\ell}_m = \left\{ \bx  =\left[ \bx_1^T, \cdots, \bx_{\ell}^T \right]^T : \bx_i  \in \left\{ \mathbf{0}, \be_1, \cdots, \be_m \right\}, \right. \\
  \label{eq:B} \left. \text{for every } i \in \{1, \cdots, \ell \}  \right. \Big\}.
\end{split}
\end{align}
The signal $\bX$ must take its values in $\mathcal{X}^{\ell_n}_M$.

The following theorem is a main result of the paper.

\begin{theorem}[Symmetric capacity of the Gaussian many-access channel]\label{thm:capacityKMac}
Let $n$ denote the coding blocklength, $\ell_n$ denote the total number of users, and $\alpha_n$ denote the probability a user is active, independent of other users. Suppose $\ell_n$ is nondecreasing with $n$ and
\begin{align}
\lim_{n \to \infty} \alpha_n = \alpha \in [0,1].
\end{align}
Denote the average number of active users as
\begin{align}
k_n = \alpha_n \ell_n.
\end{align}
Then the symmetric message-length capacity $B(n)$ of the Gaussian many-access channel~\eqref{eq:systemmodel}, with every user's signal-to-noise ratio (SNR) constrained by $P$, is characterized as:
\begin{enumerate}[{Case} 1)]
\item $\ell_n$ and $k_n$ are both unbounded, $k_n = O(n)$, and
\begin{align}\label{eq:cond_kn}
\ell_n e^{- \delta k_n} \to 0
\end{align}
for all $\delta >0$:  Let $\theta$ denote the limit of
\begin{align}\label{eq:gammaT}
\theta_n =  \frac{2 \ell_n H_2(\alpha_n)}{n \log (1+ k_n P)},
\end{align}
which may be $\infty$.
\begin{itemize}
  \item If $\theta <1$, then
  \begin{align} \label{eq:symmcapacity}
    B(n) = \frac{n}{2 k_n } \log(1 + k_n  P) - \frac{H_2 (\alpha_n)}{\alpha_n} .
  \end{align}
  \item If $\theta >1$, then a user cannot send even 1 bit reliably.
  \item If $\theta =1$, then the message length $\frac{\epsilon n}{2 k_n } \log(1 + k_n  P)$ is not achievable for any $\epsilon >0$.
\end{itemize}

\item $\ell_n$ is unbounded and $k_n$ is bounded:  $B(n)$ must be sublinear, i.e., the message length $\epsilon n$ is not achievable for any $\epsilon >0$.

\item $\ell_n$ is bounded, i.e., $\ell_n = \ell < \infty$ for large enough $n$: 
\begin{align}
  B(n) =
\begin{cases}
\frac{n}{2 } \log(1 + P) & \text{ if } \alpha =  0, \\
\frac{n}{2 \ell } \log(1 + \ell  P) & \text{ if } \alpha >  0.
\end{cases}
\end{align}
\end{enumerate}
\end{theorem}

A heuristic understanding of~\eqref{eq:symmcapacity} is as follows: If a genie revealed the set of the active users to the receiver, the total number of bits that can be communicated through the MnAC with $k_n$ users would be approximately $(n/2) \log(1 + k_n  P)$, hence the symmetric capacity is:
\begin{align}\label{eq:Cd}
  B_1 (n) = \frac{n}{2 k_n } \log(1 + k_n  P).
\end{align}
The total uncertainty in the activity of all $\ell_n$ users is $\ell_n H_2(\alpha_n) = k_n H_2(\alpha_n) / \alpha_n$, so the capacity penalty on each of the $k_n $ active users is $H_2 (\alpha_n) / \alpha_n$. If every user is always active, i.e., $\alpha_n = 1$, the penalty term is zero and the capacity resembles that of a multiaccess channel.

Because $\log(1+k_n P) = \log k_n + o(\log k_n)$,
the symmetric capacity~\eqref{eq:symmcapacity} can be reduced to
\begin{align}\label{eq:symmcapacity_2}
  B'(n) = \frac{n}{2 k_n } \log k_n - \frac{H_2 (\alpha_n)}{\alpha_n} .
\end{align}
We prefer the form of~\eqref{eq:symmcapacity} for its connection to the original capacity formula for the Gaussian multiaccess channel.

\begin{figure}
  \centering
  \includegraphics[width=\columnwidth]{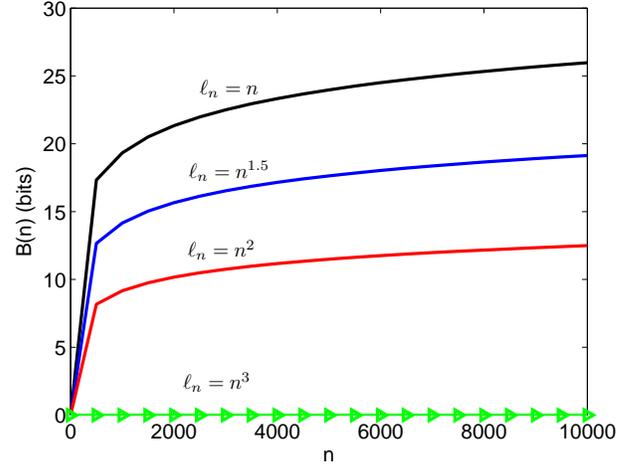}\\
  \caption{Plot of $B(n)$ given by~\eqref{eq:symmcapacity}, where $P=10$, $k_n = n/4$.}\label{fig:Cn}
\end{figure}

Fig.~\ref{fig:Cn} illustrates the message length $B(n)$ given
by~\eqref{eq:symmcapacity} with $P=10$ (i.e., the SNR is 10
dB), $k_n = n/4$, and different scalings of the user number $\ell_n $.
Evidently, $B(n)$ is sub-linear in $n$, and it depends on the scaling of $k_n$ and $\ell_n$, whose effects cannot be captured by the conventional multiaccess channel capacity result.  In particular, if $\ell_n$ grows too quickly (e.g., $\ell_n=n^3$), a typical user cannot transmit a single bit reliably. 

We have the following result on the ``overhead factor'' $\theta_n$, which is easily established by letting $n\to\infty$ in~\eqref{eq:gammaT}:

\begin{proposition}\label{prop:calgammaT}
	Let $\theta_n$ be defined as in~\eqref{eq:gammaT}. Consider the regime $k_n = \Theta(n)$. The following holds as $n \to \infty$:
	\begin{enumerate}[{Case} 1)]
		\item If $\ell_n  = \lceil a n \rceil$ for some constant $a >0$, then $\lim_{n\to\infty} \theta_n = 0$.
		\item If $\ell_n = \lceil a n^d \rceil$ for some constant $a >0$, $d >1$ and $c = \lim_{n \to \infty } k_n/n$, then $\theta_n \to 2 c (d-1) $.
	\end{enumerate}
\end{proposition}

Proposition~\ref{prop:calgammaT} demonstrates the overhead of active user identification as a function of the growth rate of $\ell_n$. When $\ell_n$ grows linearly in $n$, the cost of detecting the set of active users is negligible when amortized over a large number of channel uses. On the other hand, when $\ell_n$ grows too quickly in $n$, $\theta_n$ could be larger than 1. For user identification not to use up all channel uses, we need
\begin{align}
d< 1 + \frac{1}{2} \limsup_{n \to \infty} \frac{n}{k_n}.
\end{align}

In Theorem~\ref{thm:capacityKMac}, the assumptions in Case 1 preclude two uninteresting sub-cases:
i) the total number of users $\ell_n$ grows exponentially in $n$; and
ii) the average number of active users $k_n$ grows faster than linear in the blocklength $n$.
For example, if $k_n=n (\log n)^2$, a typical user will not be able to transmit a single bit reliably as $n$ increases to infinity.

Time sharing with power allocation, which can achieve the capacity of the conventional multiaccess channel~\cite{cover2006elements}, is inadequate for the MnAC in general. For example, if $k_n = 2n$, it cannot be guaranteed that every active user has at least one channel use.  Moreover, if $k_n = n$ and each user applies all energy in a single exclusive channel use, the resulting data rate is generally poor.

\subsection{The User Identification Cost}

As a by-product in the proof of Theorem~\ref{thm:capacityKMac}, we can derive the fundamental limits of user identification (without data transmission), where every user is active with certain probability and the receiver aims to detect the set of active users. To quantify the cost of user identification, we denote the total number of users as $\ell$ and let other parameters depend on $\ell$. (This is in contrast to the setting in Section~\ref{sec:message_capacity}.) The probability of a user being active is denoted as $\alpha_{\ell}$, and the average number of active users is denoted as $k_{\ell} = \alpha_{\ell} \ell$. Suppose $n_0$ symbols are used for user identification purpose. Let $\bX^a \in \mathbb{R}^{\ell}$ be a random vector, which consists of independent and identically distributed (i.i.d.) Bernoulli entries with mean $\alpha_{\ell}$. Then the received signal is
\begin{align}\label{eq:identify}
\bY^a = \bSS^a \bX^a + \bZ^a,
\end{align}
where $\bZ^a$ consists of $n_0$ i.i.d. standard Gaussian entries, and $\bSS^a = [\bS^a_1 \cdots, \bS^a_{\ell}]$ with $\bS^a_j \in \mathbb{R}^{n_0}$ being the signature of user $j$. Moreover, the realization of the signature must satisfy the following power constraint:
\begin{align}\label{eq:pow_constraint_signature}
\frac{1}{n_0} \sum_{i = 1}^{n_0} (\bs^a_{ki})^2 \leq P.
\end{align}

\begin{definition}[Minimum user identification cost]
We say the identification is erroneous in case of any miss or false alarm.
For the channel described by~\eqref{eq:identify}, the minimum user identification cost is said to be $n(\ell)$ if $n(\ell)>0$ and for every $0 < \epsilon <1$,
there exists a signature code of length $n_0 = (1+\epsilon) n(\ell)$ such that the probability of erroneous identification vanishes as $\ell \to \infty$,
whereas the error probability is strictly bounded away from zero if $n_0 = (1-\epsilon) n(\ell)$.
\end{definition}

As in the case of capacity, the definition focuses on the asymptotics, so the minimum cost function $n (\cdot)$ is not unique.
The random user identification problem has been studied in the context of compressed sensing problem~\cite{wainwright2009information,aksoylar2013sparse}. The following theorem gives a sharp characterization of how many channel uses $n_0$ are needed for reliable identification.

\begin{theorem}[Minimum identification cost through the Gaussian many-access channel]\label{thm:capacity_identify}
Let the total number of users be $\ell$, where each user is active with the same probability. Suppose the average number of active users $k_{\ell}$ satisfies
\begin{align}\label{eq:assump_identify}
\lim_{\ell \to \infty} \ell e^{-\delta k_{\ell}} = 0
\end{align}
for all $\delta >0$.
Let
\begin{align}\label{eq:ident_capacity_1}
n(\ell) = \frac{ \ell H_2 (k_{\ell} / \ell) }{ \frac{1}{2} \log(1+ k_{\ell} P) }.
\end{align}
The asymptotic identification cost is characterized as follows:
\begin{enumerate}[{Case} 1)]
\item As $
k_{\ell} \to \infty$,
$n(\ell) / k_{\ell}$ 
converges to a strictly positive number or goes to $+\infty$: The minimum user identification cost is  $n (\ell)$.

\item $\lim_{k_{\ell} \to \infty} n(\ell) / k_{\ell} =0$: A signature length of $n_0 = \epsilon k_{\ell}$ yields vanishing error probability for any $\epsilon >0$; on the other hand, if $n_0 \leq (1-\epsilon) n (\ell)$, then the identification error cannot vanish as $\ell \to \infty$.
\end{enumerate}
\end{theorem}

\begin{figure}
  \centering
  \includegraphics[width=\columnwidth]{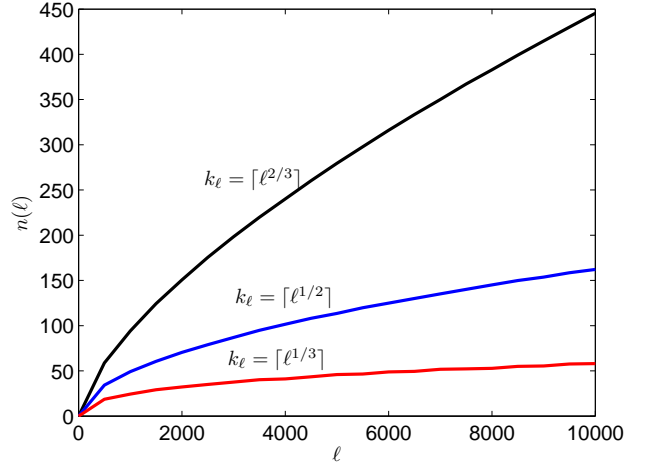}\\
  \caption{Plot of $n(\ell)$ specified in Theorem~\ref{thm:capacity_identify}, where $P=10$, i.e., SNR = 10 dB.}\label{fig:capacity_identify}
\end{figure}

Note that~\eqref{eq:assump_identify} implies $k_{\ell} \to \infty$ as $\ell \to \infty$. In the special case where $k_{\ell} = \lceil {\ell}^{1/d} \rceil$ for some $d > 1$, the minimum user identification cost is $n(\ell) = 2(d-1) k_{\ell} + o(k_{\ell})$, which is linear in the number of active users. The minimum cost function $n(\ell)$ is illustrated in Fig.~\ref{fig:capacity_identify}.

In Sections~\ref{sec:converse}--\ref{sec:achievability}, we first prove the converse part of Theorem~\ref{thm:capacityKMac}, which can be particularized to prove the converse part of Theorem~\ref{thm:capacity_identify}.  We then prove the achievability part of Theorem~\ref{thm:capacity_identify}, which is a crucial step leading to the achievability part of Theorem~\ref{thm:capacityKMac}. 

\section{Converse of Theorem~\ref{thm:capacityKMac} (MnAC Capacity)}
\label{sec:converse}

In this section, we prove the converse for the three cases in Theorem~\ref{thm:capacityKMac}.

\subsection{Converse for Case 1: unbounded $\ell_n$ and unbounded $k_n$}

This proof requires more work than a straightforward use of Fano's inequality, because the size of the joint input alphabet may increase rapidly with the blocklength. To overcome this difficulty, define for every given $\delta \in (0,1)$,
\begin{align}\label{eq:B1}
\Bcal_{m}^{\ell} (\delta, k) = \left\{ \bx \in \mathcal{X}_m^{\ell}:  1 \leq \| \bx \|_0 \leq (1+ \delta) k \right\},
\end{align}
which can be thought of as an $\ell_0$ ball but the origin.
Since $\bX$ in~\eqref{eq:systemmodel2} is a binary vector, whose expected support size is $k_n$, it is found in $\Bcal_{M}^{\ell_n} (\delta, k_n)$ with high probability for large $n$.

Based on the input distribution described in Section~\ref{sec:systemmodel},
\begin{align}\label{eq:lbHB}
H(\bX) & = \ell_n H(\bX_1) = \ell_n (H_2(\alpha_n) + \alpha_n \log M) .
\end{align}
Let $E = 1 \{ \hat{\bX} \neq \bX \} $ indicate whether the receiver makes an error, where $\hat{\bX}$ is the estimation of $\bX$. Consider an $(M,n)$ code satisfying the power constraint~\eqref{eq:powerconst} with $\prob_e^{(n)} = \prob\{E = 1\}$. The input entropy $H(\bX)$ can be calculated as
\begin{align}
& H(\bX) = H(\bX | \bY) + I(\bX; \bY) \\
& \quad = H\left(\bX, 1\left\{ \bX \in \Bcal_M^{\ell_n} (\delta, k_n) \right\} | \bY \right) + I(\bX; \bY) \\
\nonumber & \quad =  H\left( 1\left\{ \bX \in\Bcal_M^{\ell_n} (\delta, k_n) \right\} | \bY \right) + \\
\label{eq:HbX} & \qquad H\left(\bX | 1\left\{ \bX \in \Bcal_M^{\ell_n} (\delta, k_n) \right\}, \bY \right) + I(\bX; \bY)  ,
\end{align}
where we used the chain rule of the entropy to obtain~\eqref{eq:HbX}. Because the error indicator $E$ is determined by $\bX$ and $\bY$, we can further obtain
\begin{align}
\nonumber & H(\bX)  = H \left( 1 \left\{ \bX \in \Bcal_M^{\ell_n} (\delta, k_n) \right\}  | \bY  \right) +  \\
& \qquad  H \left( \bX , E| \bY, 1 \left\{ \bX \in \Bcal_M^{\ell_n} (\delta, k_n) \right\}  \right) + I(\bX; \bY) \\
\nonumber & =  H \left( 1 \left\{ \bX \in \Bcal_M^{\ell_n} (\delta, k_n) \right\}  | \bY \right) + \\
\nonumber & \qquad H\left(  E|\bY , 1 \left\{ \bX \in \Bcal_M^{\ell_n} (\delta, k_n) \right\} \right )  + \\
&  \qquad H\left( \bX | E, \bY, 1 \left\{ \bX \in \Bcal_M^{\ell_n} (\delta, k_n) \right\} \right) + I(\bX; \bY) \\
\nonumber & \leq H_2 \left( P \left\{\bX \in \Bcal_M^{\ell_n} (\delta, k_n) \right\} \right) + H_2 \left(\prob_e^{(n)}\right) + \\
&  \quad H \left( \bX | E, \bY, 1 \left\{ \bX \in \Bcal_M^{\ell_n} (\delta, k_n) \right\} \right) + I(\bX; \bY) \\
\nonumber &  \leq 2 \log 2 + H \left( \bX | E, \bY, 1 \left\{ \bX \in \Bcal_M^{\ell_n} (\delta, k_n) \right\} \right) \\
\label{eq:Fanoub}& \qquad + I(\bX; \bY).
\end{align}

In the following, we will upper bound $I(\bX ; \bY)$ and $H \left( \bX | E, \bY, 1 \left\{ \bX \in \Bcal_M^{\ell_n} (\delta, k_n) \right\}  \right)$.
\begin{lemma}\label{lemma:hy}
	Suppose $\bX$ and $\bY$ follow the distribution described by~\eqref{eq:systemmodel2}, then
	\begin{align}
	I(\bX ; \bY) \leq \frac{n}{2} \log \left(1 + k_n P \right).
	\end{align}
\end{lemma}

\begin{IEEEproof}
	See Appendix~\ref{append:Lemmahy}.
\end{IEEEproof}

\begin{lemma}\label{lemma:HBE}
	Suppose $\bX$ and $\bY$ follow the distribution described by~\eqref{eq:systemmodel2}. If $k_n$ is an unbounded sequence satisfying~\eqref{eq:cond_kn},  then for large enough $n$,
\begin{align}
  \begin{split}
    H & \left( \bX | E, \bY, 1 \left\{ \bX \in \Bcal_M^{\ell_n} (\delta, k_n) \right\} \right) \leq \\
    & 4 \prob_e^{(n)} \left( k_n \log M +  k_n + \ell_n H_2 ( \alpha_n) \right) + \log M.
  \end{split}
\end{align}
\end{lemma}

\begin{IEEEproof}
	See Appendix~\ref{append:lemmahbe}.
\end{IEEEproof}

Combining~\eqref{eq:lbHB},~\eqref{eq:Fanoub}, and Lemmas~\ref{lemma:hy} and~\ref{lemma:HBE}, we can obtain
\begin{align}
  \begin{split}
    \ell_n H_2(\alpha_n)
    & + k_n \log M   \leq \log(4M) + \frac{n}{2} \log (1+ k_n P)  \\
    & + 4 \prob_e^{(n)} (k_n \log M +  k_n + \ell_n H_2(\alpha_n) )  .
  \label{eq:ub_conv}
\end{split}
\end{align}
Dividing both sides of~\eqref{eq:ub_conv} by $k_n$ and rearranging the terms, we have
\begin{align}
\nonumber & \left( 1 - 4  \prob_e^{(n)}   \right) \log M - \frac{1}{k_n} \log M + \left( 1 - 4  \prob_e^{(n)}   \right)  \frac{ H_2 ( \alpha_n)}{\alpha_n} \\
&\qquad \leq B_1 (n)  + \frac{\log 4}{k_n} + 4 \prob_e^{(n)} ,
\end{align}
where $B_1 (n)$ is defined as~\eqref{eq:Cd}.
Since $k_n \to \infty$, we have for large enough $n$,
\begin{align}
  \begin{split}
    \left( 1 - 4  \prob_e^{(n)}  - \frac{1}{k_n} \right)
    \left( \log M + \frac{ H_2 ( \alpha_n)}{\alpha_n} \right)  \quad  \\
\label{eq:ublogM1} \quad \leq B_1 (n) + \delta + 4 \prob_e^{(n)} .
  \end{split}
\end{align}

Since $\prob_e^{(n)} $ vanishes and $k_n \to \infty$ as $n$ increases and $\delta$ can be chosen arbitrarily small, according to~\eqref{eq:ublogM1}, given any $\epsilon >0$, there exists some $\delta$ and for large enough $n$ such that
\begin{align}
\label{eq:ublogM_0} \log M & \leq (1 + \epsilon) B_1(n) - \frac{H_2(\alpha_n)}{\alpha_n} \\
\label{eq:ublogM_1}& = (1 + \epsilon - \theta_n) B_1(n) ,
\end{align}
where $\theta_n$ is defined as~\eqref{eq:gammaT}, whose limit is denoted as $\theta$.
Since~\eqref{eq:ublogM_1} holds for arbitrary $\epsilon$, if $\theta >1$, there exists a small enough $\epsilon$ such that $\log M < 0$ for large enough $n$. It implies $B(n)=0$, meaning that an average user cannot send a single bit of information reliably. If $\theta = 1$, then~\eqref{eq:ublogM_1} implies that for large enough $n$, $\log M < \epsilon B_1(n)$ for any $\epsilon>0$.

If $\theta <1$, $B(n)$ given by~\eqref{eq:symmcapacity} can be written as
\begin{align}\label{eq:capacity}
B(n) = (1 - \theta_n) B_1(n).
\end{align}
The message length can be further upper bounded as
\begin{align}
\label{eq:ublogM} \log M \leq \left( 1 + \frac{\epsilon}{1 - \theta_n} \right) B(n) ,
\end{align}
which implies $\log M \leq (1+\epsilon) B(n)$ for any arbitrarily small $\epsilon$. Thus, the converse for Case 1 is established.

\subsection{Converse for Case 2: unbounded $\ell_n$ and bounded $k_n$}

The converse claim is basically that no linear growth in message length is achievable. Suppose that, to the contrary, $\limsup_{n \to \infty} B(n) / n = C$ for some $C>0$. There must exist some $k_0 \geq 1$ such that $\frac{1}{2 k_0} \log (1+ k_0 P) < C$. Then $C$ is at least the symmetric capacity of the conventional multiaccess channel with $k_0$ users. However, as $n \to \infty$, there is a nonvanishing probability that the number of active users is greater than $2 k_0$. In this case, from the result of conventional multiaccess channel capacity, the symmetric capacity must be no greater than $\frac{1}{4 k_0} \log (1+2 k_0 P)$.  Letting each user transmit a message length of $B(n)$, which is greater than $\frac{1}{4 k_0} \log (1+ 2 k_0 P)$, would yield a strictly positive error probability. Hence the converse is proved.

\subsection{Converse for Case 3: bounded $\ell_n$}

If $\alpha_n \to 0$, a transmitting user sees no interference with probability $(1 - \alpha_n)^{\ell_n-1} \to 1$. The converse is obvious because $\frac{1}{2} \log(1+P) $ is the conventional capacity for the point-to-point channel. The achievable message length cannot exceed $\frac{n}{2} \log(1+P) $ asymptotically.

If $\alpha_n \to \alpha >0$, the number of active users is a binomial random variable. (The channel is nonergodic.) The probability that all $\ell$ users are active is $\alpha^{\ell}>0$. Hence the converse follows from the symmetric rate $\frac{1}{2 \ell} \log (1 + \ell P)$ for the conventional multiaccess channel with $\ell$ users.

\section{Proof of Theorem~\ref{thm:capacity_identify} (the Identification Cost)}
\label{sec:proof_capacity_identify}

In this section, we prove the converse and achievability of the minimum user identification cost.

\subsection{Converse of Theorem~\ref{thm:capacity_identify}}

In either of the two cases in Theorem~\ref{thm:capacity_identify}, it suffices to show that the probability of error cannot vanish if $n_0 = (1 - \epsilon) n(\ell)$ for any $0 < \epsilon <1$. The converse of Theorem~\ref{thm:capacity_identify} follows exactly from that of Theorem~\ref{thm:capacityKMac} by replacing $M = 1$ and letting $n = n_0$. According to~\eqref{eq:ublogM_0}, in order to achieve vanishing error probability for random user identification, for any $0 < \epsilon <1$,
\begin{align}
(1+\epsilon) \frac{n_0}{2 k_{\ell}} \log (1 + k_{\ell} P) \geq \frac{H_2 (\alpha_{\ell}) }{\alpha_{\ell}}.
\end{align}
Therefore, the length of the signature must satisfy
\begin{align}
n_0 > (1- \epsilon) \frac{\ell H_2(\alpha_{\ell})}{ \frac{1}{2} \log (1+ k_{\ell} P)}
\end{align}
for sufficiently large $\ell$.

\subsection{Achievability of Theorem~\ref{thm:capacity_identify}}

Let $n (\ell)$ be given by~\eqref{eq:ident_capacity_1}. Pick an arbitrary fixed $\epsilon \in (0, P)$. In the following, we will show that we can achieve vanishing error probability in identification using signature length
\begin{align}\label{eq:ident_n0}
  n_0 =
\begin{cases}
\left( 1+ \epsilon \right) n (\ell), & \text{ if  }  \lim \limits_{k_{\ell} \to \infty} n (\ell) / k_{\ell} >0 \\
\epsilon k_{\ell}, & \text{ if  } \lim\limits_{k_{\ell} \to \infty} n (\ell) / k_{\ell} = 0 .
\end{cases}
\end{align}

We provide a user identification scheme whose error probability is upper bounded by $e^{- c k_{\ell} }$ for some positive constant $c$ dependent on $\epsilon$. Let the signatures of each user $\bS_k^a$ be generated according to i.i.d. Gaussian distribution with zero mean and variance
\begin{align}
P'= P - \epsilon.
\end{align}
The receiver seeks a binary activity vector, whose weight does not exceed the average $k_{\ell}$ by a small fraction, that best explains the received signal.
This is formulated as an optimization problem:
\begin{subequations}\label{eq:decode}
\begin{align}
  \minimize \quad
  & \| \bY^a - \bSS^a \bx \|_2^2 \\
  \subjectto \quad
  & \bx \in \{0,1\}^{\ell} \\
  & \sum_{i = 1}^{\ell} x_i \leq (1 + \delta_{\ell}) k_{\ell}  ,
\end{align}
\end{subequations}
where $\delta_{\ell}$ controls the maximum weight.
We choose $\delta_{\ell} $ to be some monotone decreasing sequence such that $\delta_{\ell}^2 k_{\ell}$ increases unboundedly and $\delta_{\ell} \log k_{\ell}  \to 0$. Specifically, we let
\begin{align}\label{eq:deltal}
\delta_{\ell} = k_{\ell}^{- \frac{1}{3}}.
\end{align}

Denote $\Ecal_d$ as the event of detection error and $\mathcal{F}_{j}$ as the event that the signature of the $j$-th user violates the power constraint~\eqref{eq:pow_constraint_signature}, $j = 1, \cdots, \ell$.  The identification error probability $\prob_{e}^{(\ell)}$ is upper bounded as
\begin{align}
  \prob_{e}^{(\ell)}
  &\le \prob \left\{\Ecal_d \cup \left( \cup^\ell_{j=1} \mathcal{F}_j \right) \right\} \\
  &\le \prob \left\{\Ecal_d  \right\} + \ell \prob \left\{ \mathcal{F}_{1} \right\}
\end{align}
using the union bound and the fact that all codewords are identically distributed.


Furthermore,
\begin{align}
 \ell \prob \left\{ \mathcal{F}_{1} \right\}& =  \ell \prob \left\{ \sum_{i=1}^{n_0} (S^{a }_{1i})^2 > n_0 P \right\} \\
\label{eq:probpow} &\leq \ell e^{ - c n_0},
\end{align}
where $c$ is some positive number (which depends on $\epsilon$) due to large deviation theory for the sum of i.i.d. Gaussian random variables~\cite{durrett2010probability}. In the first case of~\eqref{eq:ident_n0}, since $\lim_{k_{\ell} \to \infty} n(\ell) / k_{\ell} >0$, there exists some $a>0$ such that for large enough $\ell$, $n_0 \geq (1+\epsilon) a k_{\ell}$. It means that in either case of~\eqref{eq:ident_n0}, $n_0 {\geq_{\ell}} \min \left( (1+\epsilon) a,\epsilon \right) k_{\ell}$, so~\eqref{eq:probpow} implies
\begin{align}
\ell \prob \left\{ \mathcal{F}_{1} \right\} \leq_{\ell} \ell e^{- \delta k_{\ell}}
\end{align}
for some $\delta >0$, which vanishes as $\ell \to \infty$ by assumption~\eqref{eq:assump_identify}.

We next derive an upper bound of the probability of detection error $\prob \left\{\Ecal_d  \right\}$. Clearly,
\begin{align}
\prob \{ \Ecal_d \} &= \Exp \left\{ \prob \{\Ecal_d | \bX^a \} \right\} \\
\nonumber & \leq \prob \{\bX^{a} \notin  \Bcal_{1}^{\ell} (\delta_{\ell} , k_{\ell}) \} + \\
\label{eq:PeStage1Error} & \quad \sum_{\bx \in \Bcal_{1}^{\ell} (\delta_{\ell} , k_{\ell}) }   \prob \{\Ecal_d | \bX^a = \bx \} \prob \left\{ \bX^a = \bx \right\} .
\end{align}
The support size of the transmitted signal $\bX^a$ as defined in~\eqref{eq:identify} follows the binomial distribution ${\rm Bin} (\ell, k_{\ell}/ \ell)$. By the Chernoff bound for binomial distribution~\cite{arratia1989tutorial},
\begin{align}
 \nonumber & \prob \{\bX^a \notin \Bcal_{1}^{\ell} (\delta_{\ell} , k_{\ell}) \}  \\
 &= \prob \left\{  \sum_{i=1}^{\ell} X_i^a   > (1+\delta_{\ell} ) k_{\ell} \right\} +  \prob \left\{  \sum_{i=1}^{\ell} X_i^a  = 0 \right\}  \\
\label{eq:ubPB2} &\leq  \exp \left( - k_{\ell} \delta_{\ell}^2 /3 \right) + (1- k_{\ell} / \ell)^{\ell},
\end{align}
which vanishes due to~\eqref{eq:deltal} and the fact that $(1- k_{\ell} / \ell)^{\ell}$ vanishes for unbounded $k_{\ell}$.
In other words, the number of active user is smaller than $(1 + \delta_{\ell}) k_{\ell}$ with high probability.
In order to prove Theorem~\ref{thm:capacity_identify}, it suffices to show that the second term on the right-hand side (RHS) of~\eqref{eq:PeStage1Error} vanishes.

Pick arbitrary $\bx^{\ast} \in \Bcal_{1}^{\ell} (\delta_{\ell}, k_{\ell})$. Let its support be $A^{\ast}$, which must satisfy $1 \leq |A^{\ast}| \leq (1 + \delta_{\ell}) k_{\ell}$. We write $\prob \{\Ecal_d | \bX^a = \bx^{\ast} \}$ interchangeably with $\prob \{\Ecal_d | A^{\ast} \}$, because there is a one-to-one mapping between $\bx^{\ast}$ and $A^{\ast}$. In the remainder of this subsection, we analyze the decoding error probability conditioned on a fixed $A^{\ast}$ and drop the conditioning on $A^{\ast}$ for notational convenience, i.e., $P \left\{ \Ecal_{d}  \right\}$ implicitly means $\prob \left\{ \Ecal_{d} | A^{\ast} \right\}$. The randomness lies in the signatures $\bSS^a$ and the received signal $\bY^a$ from $\bx^{\ast}$. Define
\begin{align}\label{eq:Ta}
  T_A
  = \bigg\| \bY^a - \sum_{i \in A} \bS^a_i \bigg\|^2
  - \bigg\| \bY^a - \sum_{i \in A^{\ast}} \bS^a_i \bigg\|_2^2.
\end{align}
According to the decoding rule~\eqref{eq:decode}, a detection error may occur only if there is some $A \subseteq \{1, \cdots, \ell \}$ such that $A \neq A^{\ast} $, such that $ |A| \leq (1 + \delta_{\ell}) k_{\ell} $, and $ T_A \leq 0$. Hence,
\begin{align}\label{eq:setEd1}
\Ecal_d  =  \bigcup_{ \substack{ A \subseteq \{1, \cdots, \ell\}: \\|A| \leq (1+ \delta_{\ell}) k_{\ell}, A \neq A^{\ast} } } \{ T_A \leq 0 \}.
\end{align}

\begin{figure}
  \centering
  \small
\begin{tikzpicture} 
  \draw (-1.5,2.25) node {$A^{\ast}$};
  \draw (1.5,2.25) node {$A$};
  \draw (-1.25,1.6) ellipse [x radius=2, y radius=0.45] node (deltaone) {$\quad A_1=A^{\ast}\setminus A\qquad A\cap A^{\ast}$};
  \draw (1.25,1.6) ellipse [x radius=2, y radius=0.45] node {$\qquad\quad A_2=A\setminus A^{\ast}$};
\end{tikzpicture}
  \caption{The set relationship.}
  \label{fig:LocalPattern}
\end{figure}
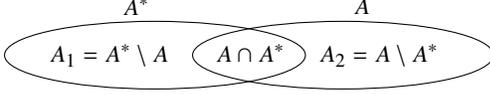

In the following, we divide the exponential number of error events in~\eqref{eq:setEd1} into a relatively small number of classes. We will show that the probability of error of each class vanishes and so does the overall error probability. Specifically, we write the union over $A$ according to the cardinality of the sets $A^{\ast} \cap A$ and $A \backslash A^{\ast}$. Let $w_1 = |A_1|$ and $ w_2 = |A_2| $, where $A_1 = A^{\ast} \backslash A$ represents the set of misses and $A_2 = A \backslash A^{\ast}$ represents the set of false alarms.  (The set relationship is depicted by Fig.~\ref{fig:LocalPattern}.)  Then $(w_1,w_2)$ must satisfy $w_1 \leq |A^{\ast}|$, $w_2 \leq |A|$, and $|A^{\ast}| +w_2 = |A| + w_1$. According to the decoding rule~\eqref{eq:decode}, $(w_1,w_2)$ must be found in the following set:
\begin{align}
  \begin{split}
 & \mathcal{W}^{(\ell)} = \left\{ (w_1, w_2): w_1 \in \{ 0, 1, \cdots, |A^{\ast}| \},  \right. \\
 & \qquad \left. w_2 \in \{ 0,1, \cdots, (1+  \delta_{\ell}) k_{\ell} \}, \right. \\
\label{eq:setW} & \qquad \left. w_1 + w_2 >0, |A^{\ast}|  + w_2 \leq (1+  \delta_{\ell}) k_{\ell} + w_1 \right\}.
  \end{split}
\end{align}

We further define the event $\Ecal_{w_1, w_2}$ as
\begin{align}\label{eq:setEw}
\Ecal_{w_1, w_2}  = \bigcup_{ \substack { A \subseteq \{1, \cdots, \ell\}: \\ |A^{\ast} \backslash A| =w_1, |A \backslash A^{\ast}| = w_2} } \{ T_A \leq 0 \}.
\end{align}
By~\eqref{eq:setEd1}, $\Ecal_d  \subseteq \cup_{(w_1,w_2) \in \mathcal{W}^{(\ell)}  } \Ecal_{w_1, w_2}$. Hence \begin{align}
\prob \{\Ecal_d \} \leq \sum_{(w_1,w_2) \in \mathcal{W}^{(\ell)}  } \prob \{ \Ecal_{w_1, w_2}\}.
\end{align}
We will show that when $\ell$ is large enough, there exists some constant $c_0 >0$ such that $\prob \{ \Ecal_{w_1, w_2} \} \leq e^{- k_{\ell} c_0}$ for all $(w_1, w_2) \in \mathcal{W}^{(\ell)} $.

Define
\begin{align}
\mathcal{A}_1 (w_1) = \left\{A_1: A_1 \subseteq A^{\ast}, |A_1| =  w_1 \right\}
\end{align}
and
\begin{align}
\mathcal{A}_2(w_2) = \left\{A_2: A_2 \subseteq \{1, \cdots, \ell \} \backslash A^{\ast}, |A_2| = w_2 \right\}.
\end{align}
Then any $A$ leading to an error event in $\Ecal_{w_1, w_2}$ specified by~\eqref{eq:setEw} can be written as $A = A_2 \cup (A^{\ast} \backslash A_1)$, for some $A_1 \in \mathcal{A}_1 (w_1)$ and $A_2 \in \mathcal{A}_2 (w_2)$. Therefore,~\eqref{eq:setEw} gives
\begin{align}
\Ecal_{w_1, w_2} = \bigcup_{A_1 \in \mathcal{A}_1(w_1) } \bigcup_{ A_2 \in  \mathcal{A}_2(w_2) } \{ T_A \leq 0 \},
\end{align}
which implies
\begin{align}\label{eq:1Ew}
1 \left\{ \Ecal_{w_1, w_2} \right\} \leq \sum_{A_1 \in \mathcal{A}_1(w_1)} \left( \sum_{A_2 \in \mathcal{A}_2(w_2)} 1 \left\{ T_A \leq 0 \right\} \right)^{\rho}
\end{align}
for all $\rho \in [0,1]$. As a result,
\begin{align}
  &\prob \left\{ \Ecal_{w_1, w_2}  \right\}
  = \Exp \left\{ 1 \left\{ \Ecal_{w_1, w_2} \right\} \right\} \\
  &\qquad \leq \sum_{A_1 \in \mathcal{A}_1(w_1)} \Exp \left\{ \left( \sum_{A_2 \in \mathcal{A}_2(w_2)} 1 \left\{ T_A \leq 0 \right\} \right)^{\rho} \right\}
\label{eq:PEw}
\end{align}
where the expectation is taken over $(\bSS^a, \bY^a)$. We further calculate the expectation by first conditioning on $(\bSS^a_{A^{\ast}}, \bY^a)$ as follows:
\begin{align}
  \nonumber
  & \prob \left\{ \Ecal_{w_1, w_2}  \right\} \\
  & \leq \sum_{A_1 \in \mathcal{A}_1(w_1)}  
  \!\!\! \Exp \Bigg\{ \Exp \Bigg\{
  \bigg( \sum_{A_2 \in \mathcal{A}_2(w_2)}
  \!\!\! 1 \left\{ T_A \leq 0 \right\} \bigg)^{\rho}
  \bigg| \bSS^a_{A^{\ast}}, \bY^a
  \Bigg\} \Bigg\} \\
  & \leq \sum_{A_1 \in \mathcal{A}_1(w_1)}
  \!\!\! \Exp \Bigg\{ \Bigg[ \Exp \Bigg\{
  \sum_{A_2 \in \mathcal{A}_2(w_2)}
  \!\!\! 1 \left\{ T_A \leq 0 \right\}  \bigg| \bSS^a_{A^{\ast}}, \bY^a \Bigg\}
  \Bigg]^{\rho} \Bigg\} ,
 \label{eq:1EwStep2}
\end{align}
where the expectation is taken first with respect to the probability measure $p_{\bSS^a_{ \{1, \cdots, \ell \} \backslash A^{\ast} } | \bSS^a_{A^{\ast}}, \bY^a}$ and then with respect to the probability measure $p_{\bSS^a_{A^{\ast}}, \bY^a}$; and Jensen's inequality is applied in~\eqref{eq:1EwStep2} to the concave function $x^{\rho}$, $0 < \rho \leq 1$. Since $\bSS^a_{ \{1, \cdots, \ell \} \backslash A^{\ast} } $ and $ \bSS^a_{A^{\ast}}$ are independent and $\bY^a$ only depends on $\bSS^a_{A^{\ast}}$, we have $p_{\bSS^a_{ \{1, \cdots, \ell \} \backslash A^{\ast} } | \bSS^a_{A^{\ast}}, \bY^a} (\bss_1 | \bss_2, \by) = p_{\bSS^a_{ \{1, \cdots, \ell \} \backslash A^{\ast} } } (\bss_1)$.
The inner expectation in~\eqref{eq:1EwStep2} is taken with respect to the probability measure $p_{\bSS^a_{A_2}  }$ for each $A_2 \in \mathcal{A}_2(w_2)$. Since the entries of $\bSS^a$ are i.i.d., the inner expectation yields identical results for all $A_2 \in \mathcal{A}_2(w_2)$ and the outer expectation yields identical results for all $A_1 \in \mathcal{A}_1(w_1)$.

The number of choices for $A_1$ is $\binom{|A^{\ast}|}{ w_1}$, whereas the number of choices for $A_2$ is no greater than $\binom{\ell}{w_2}$. Therefore, we apply the union bound to obtain
\begin{align}
\nonumber \prob \left\{ \Ecal_{w_1, w_2} \right\} & \leq \binom{|A^{\ast}|}{ w_1} \binom{\ell}{w_2}^{\rho}  \times \\
\label{eq:1EwStep4}& \quad \Exp \left\{  \left[ \Exp \left\{  1 \left\{ T_{A} \leq 0 \right\} \big| \bSS^a_{A^{\ast}}, \bY^a   \right\} \right]^{\rho} \right\},
\end{align}
where $A$ is now a fixed representative choice with $|A^{\ast} \backslash A| = w_1$ and $|A \backslash A^{\ast}| = w_2$.

We next upper bound the detection error probability by upper bounding $\Exp \left\{  1 \left\{ T_{A} \leq 0 \right\} \big| \bSS^a_{A^{\ast}}, \bY^a   \right\} $. Let
\begin{align}
p_{Y | \bS_A} ( y_i | \bs_{A,i} ) = \frac{1}{\sqrt{2 \pi}} \exp\left( - \frac{1}{2} \left( y_i - \sum_{k \in A} s_{k i} \right)^2 \right).
\end{align}
Recall that the noise entries are i.i.d. standard Gaussian.  The conditional distribution of $\by$ given that the codewords $\bss_{A}$ are transmitted is given by $p_{\bY | \bSS_A} (\by | \bss_A) = \prod_{i=1}^{ n_0} p_{Y | \bS_A} ( y_i | \bs_{A,i} )$, where $ n_0$ is the dimension of $\by$.
Then for any $\lambda \geq 0$, the following holds due to~\eqref{eq:Ta}:
\begin{align}
\nonumber & \Exp \left\{  1 \left\{ T_{A} \leq 0 \right\} \big| \bSS^a_{A^{\ast}}, \bY^a   \right\} \\
 &= \Exp \left\{  1 \left\{ \frac{p_{\bY | \bSS_A} ( \bY^a | \bSS^a_A)}{p_{\bY | \bSS_A} ( \bY^a | \bSS^a_{A^{\ast}}) }  \geq 1  \right\} \bigg| \bSS^a_{A^{\ast}}, \bY^a \right\}\\
\label{eq:Ptsub2} & \leq \Exp  \left\{ \left(  \frac{p_{\bY | \bSS_A} ( \bY^a | \bSS^a_A)}{p_{\bY | \bSS_A} ( \bY^a | \bSS^a_{A^{\ast}}) }  \right)^{\lambda}  \bigg| \bSS^a_{A^{\ast}}, \bY^a \right\} \\
\label{eq:1EwStep5}  & =   p^{-\lambda}_{\bY | \bSS_A} ( \bY^a | \bSS^a_{A^{\ast}})  \Exp  \left\{   p^{\lambda}_{\bY | \bSS_A} ( \bY^a | \bSS^a_A)     \big| \bSS^a_{A^{\ast}}, \bY^a \right\},
\end{align}
where~\eqref{eq:1EwStep5} follows because $(\bSS^a_{A^{\ast}}, \bY)$ is independent of $\bSS^a_{A_2}$. For every function $g \left(\bS_{A^{\ast}}^a , \bY^a \right)$,
\begin{align}
  \Exp \left\{ g \left(\bS_{A^{\ast}}^a , \bY^a \right) \right\} = \int_{\mathbb{R}^{n_0} } \Exp \left\{ g \left(\bS_{A^{\ast}}^a , \by \right) p_{\bY | \bSS_A} ( \by | \bSS^a_{A^{\ast}}) \right\} d \by .
\end{align}
Combining~\eqref{eq:1EwStep4} and~\eqref{eq:1EwStep5} yields
\begin{align}
\nonumber &\prob \left\{ \Ecal_{w_1, w_2}  \right\} \leq \binom{|A^{\ast}|}{ w_1} \binom{\ell}{w_2}^{\rho} \times \\
&\int_{\mathbb{R}^{n_0}}  \Exp \left\{  p^{1- \lambda \rho}_{\bY | \bSS_A} ( \by | \bSS^a_{A^{\ast}})  \left( \Exp  \left\{ p^{\lambda}_{\bY | \bSS_A} ( \by | \bSS^a_A)   \Big| \bSS^a_{A^{\ast}}   \right\} \right)^{\rho} \right\} d \by .
\end{align}

Due to the memoryless nature of the channel, i.e., $p_{\bY | \bSS_A} ( \by | \bSS^a_A) = \prod_{i=1}^{n_0} p_{Y | \bS_A} ( y_i | \bS^a_{A,i} )$, we obtain
\begin{align}
\label{eq:PEw_ub} \prob \left\{ \Ecal_{w_1, w_2}  \right\} &\leq \binom{|A^{\ast}|}{ w_1} \binom{\ell}{w_2}^{\rho}  \left( m_{\lambda, \rho} ( w_1,w_2) \right)^{n_0}
\end{align}
where
\begin{align}
\nonumber & m_{\lambda, \rho} ( w_1,w_2) = \\
\label{eq:m_lambda_rho}& \int_{\mathbb{R}}  \Exp \left\{  p^{1- \lambda \rho}_{Y | \bS_A} ( y | \bS^a_{A^{\ast}})  \left( \Exp  \left\{ p^{\lambda}_{Y | \bS_A} ( y | \bS^a_A)   \Big| \bS^a_{A^{\ast}}   \right\} \right)^{\rho} \right\} d y .
\end{align}
The product of the first two factors in the RHS of~\eqref{eq:PEw_ub} can be upper bounded as~\cite[Page 353]{cover2006elements}
\begin{align}
  \binom{|A^{\ast}|}{ w_1} \binom{\ell}{w_2}^{\rho}
  \!\!\leq \exp \left[ |A^{\ast}| H_2 \Big( \frac{w_1}{ |A^{\ast}|} \Big) + \rho \ell H_2 \left( \frac{w_2}{ \ell} \right) \right].
\end{align}
Moreover, by the Gaussian distribution of the codewords, the last factor in the RHS of~\eqref{eq:PEw_ub} can be explicitly calculated (see Appendix~\ref{append:derivehlamdarho}) to obtain
\begin{align} \label{eq:m_lambda}
  \begin{split}
    & m_{\lambda, \rho} ( w_1,w_2)  =  \exp \left(   \frac{1- \rho}{2} \log(1+ \lambda w_2 P') - \right. \\
    & \quad \left. \frac{1}{2} \log \left( 1+ \lambda (1- \lambda \rho) w_2 P'  + \lambda\rho (1-\lambda \rho) w_1 P' \right)  \right),
  \end{split}
\end{align}
where $\lambda \rho \leq 1$. Therefore, by~\eqref{eq:PEw_ub}-\eqref{eq:m_lambda},
\begin{align}\label{eq:PEw12}
\prob \{\Ecal_{w_1, w_2}  \}  \leq \exp \left(- k_{\ell} h_{\lambda,\rho}(w_1, w_2) \right),
\end{align}
where
\begin{align} \label{eq:hlambdarho}
  \begin{split}
    & h_{\lambda,\rho} (w_1, w_2) =
    - \frac{(1 - \rho) n_0}{2 k_{\ell} } \log \left( 1 + \lambda w_2 P' \right) \\
    & \;\; + \frac{ n_0}{2 k_{\ell}} \log \left( 1 + \lambda (1 - \lambda \rho) w_2 P' + \lambda \rho (1 - \lambda \rho) w_1 P' \right) \\
    & \;\;
    -  \frac{ |A^{\ast}|}{k_{\ell}} H_2\left( \frac{w_1}{  |A^{\ast}|} \right)
    - \frac{\rho \ell}{k_{\ell}} H_2 \left( \frac{w_2}{ \ell} \right).
   \end{split}
\end{align}
To show the achievability, we next show that by choosing $\lambda$ and $\rho$ properly, for large enough $\ell$, $ h_{\lambda,\rho}(w_1, w_2)  $ is strictly greater than some positive constant for all $(w_1, w_2) \in \mathcal{W}^{(\ell)}$.

\begin{lemma}\label{lemma:minhlambdarho}
Fix $\epsilon \in (0, P)$. Let $P' = P- \epsilon$. Let $n(\ell)$ be given by~\eqref{eq:ident_capacity_1} and $n_0$ be given by~\eqref{eq:ident_n0}. Suppose $n(\ell) / k_{\ell}$ has finite limit or diverges to infinity. There exists $\ell^{\ast} >0$ and $c_0>0$ such that for every $\ell \geq \ell^{\ast}$ the following holds: If the true signal $\bx^a \in \Bcal_{1}^{\ell} (\delta_{\ell} , k_{\ell})$, i.e., $1 \leq |A^{\ast}| \leq (1+\delta_{\ell}) k_{\ell}$, then for every $(w_1, w_2) \in \mathcal{W}^{(\ell)}$ with $\mathcal{W}^{(\ell)}$ defined as in~\eqref{eq:setW}, there exist $\lambda \in [0, \infty)$ and $\rho \in [0, 1]$ such that
\begin{align}\label{eq:lb_h_lambda}
h_{\lambda,\rho}(w_1, w_2) \geq c_0.
\end{align}
\end{lemma}
\begin{IEEEproof}
See Appendix~\ref{append:pflemmaminhlambdarho}.
\end{IEEEproof}

Lemma~\ref{lemma:minhlambdarho} and~\eqref{eq:PEw12} imply
\begin{align}
\prob \{\Ecal_{w_1, w_2} | A^{\ast} \}  \leq e^{- c_0 k_{\ell} },
\end{align}
for all $\ell \geq \ell^{\ast}$, $(w_1,w_2) \in \mathcal{W}^{(\ell)} $, and $ 1 \leq |A^{\ast}| \leq (1+\delta_{\ell}) k_{\ell}$. Then as long as $\ell \geq \ell^{\ast}$, for any $\bx \in \Bcal_1^{\ell} (\delta_{\ell} , k_{\ell})$,
\begin{align}
\prob \{\Ecal_d | \bX^a = \bx \} & \leq   \sum_{(w_1, w_2) \in \mathcal{W}^{(\ell)} } \prob \{ \Ecal_{w_1, w_2 } | \bX^a = \bx \} \\
& \leq  \sum_{(w_1, w_2) \in \mathcal{W}^{(\ell)} }   e^{- c_0 k_{\ell} } \\
\label{eq:ub_P_Ed}& \leq 4 k^2_{\ell} e^{- c_0 k_{\ell} },
\end{align}
where~\eqref{eq:ub_P_Ed} is due to $w_1 \leq 2 k_{\ell}$ and $ w_2 \leq 2 k_{\ell}$. Therefore, the first term in the RHS of~\eqref{eq:PeStage1Error} vanishes as $\ell$ increases. So does $\prob \{\Ecal_d \}$. Thus we can achieve arbitrarily reliable identitifcation with SNR $P'=P-\epsilon$ and signature length $n_0$ given by~\eqref{eq:ident_n0}. Since $\epsilon$ can be arbitrarily small, the achievability of Theorem~\ref{thm:capacity_identify} is established.

\section{Achievability of Theorem~\ref{thm:capacityKMac} (MnAC Capacity)}
\label{sec:achievability}

In this section, we prove the achievability part of Theorem~\ref{thm:capacityKMac} to establish the symmetric capacity of the MnAC.

\subsection{Achievability for Case 3 with bounded $\ell_n$}

As $\ell_n$ is nondecreasing, $\ell_n \to \ell$ for some constant $\ell$. If $\alpha_n \to \alpha >0$, with some positive probability all $\ell$ users are active.
Hence the achievability capacity follows from the result for the conventional multiaccess channel with $\ell$ users.

If $\alpha_n \to 0$, a transmitting user experiences a single-user channel with probability  $ (1-\alpha_n)^{\ell_n -1} \to 1$.
Therefore, it can achieve a vanishing error probability with the conventional capacity for the point-to-point channel.

\subsection{Achievability for Case 1 and Case 2 with unbounded $\ell_n$}

\begin{figure}
  \centering
  \includegraphics[width=9cm]{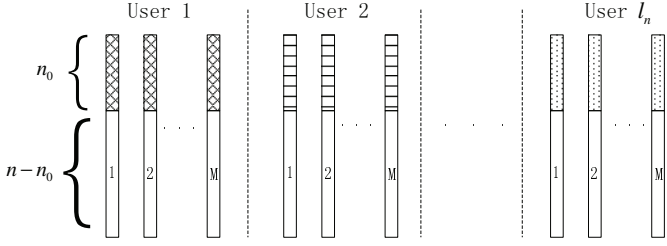}\\
  \caption{Codebook structure. Each user maintains $M$ codewords with each consisting of a message-bearing codeword prepended by a signature.}\label{fig:codebook}
\end{figure}

We first assume unbounded $k_n$ and establish the achievability result. The case of bounded $k_n$ is then straightforward.

We consider a two-stage approach: In the first stage, the set of active users are identified based on their unique signatures. In the second stage, the messages from the active users are decoded. Let $\theta_n$ and its limit $\theta$ be defined as in Theorem~\ref{thm:capacityKMac}. We consider the cases of $\theta = 0$ and $\theta > 0$ at the same time. Fix $\epsilon \in (0, \min(1,P))$. Specifically, the following scheme is used:

\begin{itemize}

\item \textit{Codebook construction:} The codebooks of the $\ell_n$ users are generated independently. Let
    \begin{align}\label{eq:NT0}
     n_0 =
     \begin{cases}
       \epsilon n , & \text{if  } \theta =  0 \\
       \left( 1 + \epsilon \right) \theta_n n, & \text{otherwise  }.
     \end{cases}
     \end{align}
     For user $k$, codeword $\bs_k(0) = \mathbf{0}$ represents silence. User $k$ also generates
     \begin{align}\label{eq:logCap1}
     M = \lceil \exp \left[ (1 - \epsilon) B(n) \right] \rceil
    \end{align}
 codewords as follows. First, generate $M$ random sequences of length $n - n_0$, each according to i.i.d. Gaussian distribution with zero mean and variance $P' = P - \epsilon$. Then generate one signature of length $n_0$ with i.i.d. $\mathcal{N} (0, P')$, denoted by $\bS_k^a$, and prepend this signature to every codeword to form $M$ codewords of length $n$.
In other words, the $w$-th codeword of user $k$ takes the shape of $\bS_k(w) = \icol{\bS_k^a \\ \bS_k^b(w)}$.
 The matrix of the concatenated codebooks of all users is illustrated in Fig.~\ref{fig:codebook}.

\item \textit{Transmission:} For user $k$ to be silent, it is equivalent to transmitting $\bs_k(0)$. Otherwise, to send message $w_k \neq 0$, user $k$ transmits $\bS_k (w_k)$.
\item \textit{Channel: } Each user is active independently with probability $\alpha_n$. The active users transmit simultaneously. The received signal is $\bY$ given by~\eqref{eq:systemmodel2}.
\item \textit{Two-stage detection and decoding:} Upon receiving $\bY$, the decoder performs the following:

1) Active user identification: Let $\bY^a$ denote the first $n_0$ entries of $\bY$, corresponding to the superimposed signatures of all active users subject to noise. $\bY^a$ is mathematically described by~\eqref{eq:identify}. The receiver detects $\bX^a$ according to~\eqref{eq:decode}. The output of this stage is a set $A \subseteq \{1, \cdots, \ell_n \}$ that contains the detected active users.


2) Message decoding: Let $\bY^b$ denote the last $n-n_0$ entries of $\bY$, corresponding to the superimposed message-bearing codewords. The receiver solves the following optimization problem:
\begin{subequations}
\begin{align}
  \minimize\quad
  & \Big\| \bY^b - \bSS^b \left[ \bx_1^T, \cdots, \bx_{\ell_n}^T \right]^T \Big\|^2 \\
  \subjectto\quad
  & \bx_k \in \mathcal{X}_M^{1}, \;k = 1, \cdots, \ell_n \\
  & \bx_k = \mathbf{0}, \quad \forall k \notin A \\
  & \bx_k \neq \mathbf{0}, \quad \forall k \in A . 
\end{align}
\end{subequations}
Basically the receiver performs the maximum likelihood decoding for the set of users in the purported active user set $A$. The position of 1 in each recovered nonzero $\bx_k$ indicates the message from user $k$.
\end{itemize}

\begin{theorem}[Achievability of the Gaussian many-access channel]\label{thm:logCap}
Let  $\theta_n$ be defined as~\eqref{eq:gammaT} and $B(n)$ be defined as~\eqref{eq:symmcapacity}. Suppose $\lim_{n \to \infty} \theta_n <1$. For the MnAC given by~\eqref{eq:systemmodel}, for any given constant $\epsilon \in (0,1)$, the message length of $(1 - \epsilon) B(n)$ is asymptotically achievable using the preceding scheme.
\end{theorem}

The remainder of this section is devoted to the proof of Theorem~\ref{thm:logCap}. In Section~\ref{subsec:pfthm_logCapStage1}, we show that the set of active users can be accurately identified in the first stage. In Section~\ref{subsec:pfthm_logCapStage2}, we show that the users' messages can be accurately decoded in second stage assuming knowledge of the active users. The results are combined in Section~\ref{subsec:achievecapacity} to establish the achievability part of Theorem~\ref{thm:logCap}.

\subsection{Optimal User Identification}
\label{subsec:pfthm_logCapStage1}

We shall invoke Theorem~\ref{thm:capacity_identify} (proved in Section~\ref{sec:proof_capacity_identify}) to quantify the cost of reliable user identification. To adapt to the notation in this section, we apply Theorem~\ref{thm:capacity_identify} with $\ell$ and $k_{\ell}$ being replaced by $\ell_n$ and $k_n$, respectively. With the change of notations, $n(\ell)$ as defined in Theorem~\ref{thm:capacity_identify} can be written as
\begin{align}
n(\ell) &= \frac{\ell_n H_2(k_n / \ell_n)}{\frac{1}{2} \log(1+k_n P)}  \\
&= \theta_n n,
\end{align}
where $\theta_n$ is given by~\eqref{eq:gammaT}.

According to Theorem~\ref{thm:capacity_identify}, choosing the signature length $n_0 = ( 1+\epsilon) \theta_n n$ and $n_0 = \epsilon k_n$ yields vanishing error probability in user identification for the case of $\lim_{n \to \infty} \theta_n n / k_n > 0$ and $\lim_{n \to \infty} \theta_n n / k_n = 0$, respectively, where $\epsilon \in (0,1)$ is an arbitrary constant. In the following, we make use of this result to prove that choosing $n_0$ according to~\eqref{eq:NT0} guarantees reliable user identification.

First, consider $\theta = 0$. By~\eqref{eq:NT0}, the signature length is $n_0 = \epsilon n$ for some $\epsilon$. In the case of $\lim_{n \to \infty} \theta_n n /k_n >0$, since $\theta_n$ vanishes, we must have $n_0 \geq_n (1+ \epsilon) \theta_n n$. In the case of $\lim_{n \to \infty} \theta_n n / k_n = 0$, since $k_n = O(n)$, $n_0 = \epsilon n$ implies $n_0 \geq_n \epsilon' k_n$ for some $\epsilon' >0$. By Theorem~\ref{thm:capacity_identify}, the choice of $n_0$ is sufficient for reliable user identification.

Second, consider $\theta >0$. By~\eqref{eq:NT0}, the signature length is $n_0 = (1+\epsilon) \theta_n n$. Since $k_n = O(n)$, we must have $\lim_{n \to \infty} \theta_n n /k_n >0$. Thus, the signature length $n_0$ obviously achieves reliable user identification by Theorem~\ref{thm:capacity_identify}.

\subsection{Achieving the MnAC Capacity with Known User Activities}
\label{subsec:pfthm_logCapStage2}


In~\cite{chen2013gaussian}, we studied a Gaussian MnAC where all users are always active and the number of users is sublinear in the blocklength, i.e., $k_n=o(n)$. In that case, random coding and Feinstein's suboptimal decoding, which achieve the capacity of conventional multiaccess channel, can also achieve the capacity of the Gaussian MnAC. Proving the achievability for faster scaling of the number of active users is much more challenging, mainly because the exponential number of possible error events prevents one from using the simple union bound.  Here, we derive the capacity of the MnAC for the case where the number of users may grow as quickly as linearly with the blocklength by lower bounding the error exponent of the error probability due to maximum-likelihood decoding.

\begin{theorem}[Capacity of the Gaussian many-access channel without random access]\label{thm:logCapStage2}
For the MnAC with $k_n$ always-active users, suppose the number of channel uses is $n$ and the number of users $k_n$ grows as $O(n)$, the symmetric capacity is
\begin{align}
B_1 (n) = \frac{n}{2 k_n } \log(1 + k_n  P).
\end{align}
In particular, for any $\epsilon \in (0,1)$, there exists a sequence of codebooks with message lengths (in nats) $B_1(n) (1 - \epsilon ) $ such that the average error probability is arbitrarily small for sufficiently large $n$.
\end{theorem}

We prove Theorem~\ref{thm:logCapStage2} in the remainder of this subsection. We can model the MnAC with known user activities using~\eqref{eq:systemmodel2} with $\alpha_n = 1$, i.e., $k_n = \ell_n$. Upon receiving the length-$n$ vector $\by$, we estimate $\bx = \left[\bx^T_1, \cdots, \bx^T_{k_n} \right]^T$ using the maximum likelihood decoding:
\begin{subequations}
\begin{align}
  \minimize \quad
  &   \| \by - \bss \bx \|^2 \\
  \subjectto \quad
  &  \bx_k = \be_m \;\; \text{for some  } m = 1, \cdots, M.
\end{align}
\end{subequations}

Define $\mathcal{F}_j$ as the event that user $j$'s codeword violates the power constraint~\eqref{eq:powerconst}, $j = 1, \cdots, k_n$. Define $\Ecal_{k}$ as the error event that $k$ users are received in error. Suppose $\prob\{\Ecal_{k} | A^{\ast} \}$ is the probability of $\Ecal_{k}$ given that the true signal is $\bx^{\ast}$ with support $A^{\ast}$. By symmetry of the codebook construction, the average error probability can be upper bounded as
\begin{align}
  \prob_e^{(n)}
  &\le \prob \left\{ \left( \cup_{k=1}^{k_n} \Ecal_k \right)
  \cup \left( \cup_{j =1}^{k_n} \mathcal{F}_j \right) \right\} \\
  \label{eq:avgPe}
  &\leq  \frac{1}{M^{k_n}} \sum_{A^{\ast}} \sum_{ k=1}^{k_n}
    \prob \{\Ecal_{k} | A^{\ast} \}
    + \sum_{j=1}^{k_n} \prob \left\{ \mathcal{F}_j \right\}.
\end{align}

Let $A$ be the support of the estimated $\bx$ according to the maximum likelihood decoding.
Define $A_1$ and $A_2$ in the same manner as that in Section~\ref{subsec:pfthm_logCapStage1}, i.e., $A_1 = A^{\ast} \backslash A$ and $A_2 = A \backslash A^{\ast}$. In this case, $|A| = |A^{\ast}| = k_n$ and $|A_2| = |A_1| = k$. Further denote $\gamma = k / k_n$ as the fraction of users subjected to errors. Then we write $\prob \{\Ecal_{k} | A^{\ast} \}$ and $\prob \{\Ecal_{\gamma} | A^{\ast} \}$ interchangeably. In the following analysis, we consider a fixed $A^{\ast}$ and drop the conditioning on $A^{\ast}$ for notational convenience. Following similar arguments leading to~\eqref{eq:PEw_ub}, letting $\lambda = \frac{1}{1 + \rho}$ and considering $\binom{k_n}{\gamma k_n}$ possible sets of $A_1$ and  $M^{\gamma k_n}$ possible sets of $A_2$, we have
\begin{align}
\nonumber \prob \{\Ecal_{\gamma}  \}  &\leq  \binom{k_n}{\gamma k_n} M^{\gamma k_n \rho}  \left( \int_{\mathbb{R}}  \Exp \left\{   p^{\frac{1}{\rho+1} }_{Y | \bS_A} ( y | \bS_{A^{\ast}})  \times  \right. \right. \\
& \qquad \left. \left. \left(  \Exp \left\{ p^{\frac{1}{\rho+1} }_{Y | \bS_A} ( y | \bS_{A})  \Big| \bS_{A^{\ast}} \right\} \right)^{\rho}      \right\} d y  \right)^n \\
\nonumber & = \binom{k_n}{\gamma k_n} M^{\gamma k_n \rho} \times \\
\nonumber & \quad  \left( \int_{\mathbb{R}}  \Exp \left\{  \Exp \left\{ p^{\frac{1}{\rho+1} }_{Y | \bS_A} ( y | \bS_{A^{\ast}})  \Big| \bS_{A^{\ast} \cap A} \right\}  \times \right. \right. \\
& \quad \left. \left. \left(  \Exp \left\{ p^{\frac{1}{\rho+1} }_{Y | \bS_A} ( y | \bS_{A})  \Big| \bS_{A^{\ast}} \right\} \right)^{\rho}      \right\} d y  \right)^n.
\end{align}
By symmetry,
\begin{align}
  \Exp \left\{ p^{\frac{1}{\rho+1} }_{Y | \bS_A} ( y | \bS_{A})   \Big| \bS_{A^{\ast}} \right\} = \Exp \left\{   p^{\frac{1}{\rho+1} }_{Y | \bS_A} ( y | \bS_{A^{\ast}})  \Big| \bS_{A^{\ast} \cap A} \right\},
\end{align}
which results in
\begin{align}
\label{eq:Peub3b} \prob \{\Ecal_{\gamma}  \}  &\leq  \binom{k_n}{\gamma k_n} M^{\gamma k_n \rho} \exp(-n E_0(\gamma, \rho) ),
\end{align}
where $E_0(\gamma, \rho)$ is defined by
\begin{align}
\nonumber & E_0(\gamma, \rho) = \\
\label{eq:E0} & -\log \left[ \int_{\mathbb{R}} \Exp \left\{ \left[  \Exp \left\{ \left(  p_{Y | \bS_A} ( y | \bS_A)  \right)^{\frac{1}{\rho+1} } \Big| \bS_{A^{\ast}} \right\} \right]^{1+\rho}  \right\} dy \right].
\end{align}
By the inequality
\begin{align}
  \binom{k_n}{\gamma k_n} \leq \exp(k_n H_2(\gamma)),
\end{align}
we can further upper bound $ \prob \{\Ecal_{\gamma}  \}$ as
\begin{align}\label{eq:Peub4}
\prob \{\Ecal_{\gamma}  \} \leq \exp \left[ - n f(\gamma, \rho) \right],
\end{align}
where
\begin{align}\label{eq:f_gamma_rho}
f(\gamma, \rho) = E_0(\gamma, \rho) - \gamma \rho \frac{k_n}{n} v(n) - \frac{k_n}{n} H_2(\gamma),
\end{align}
and $v(n) = \log M$. Intuitively, $E_0(\gamma, \rho)$ in~\eqref{eq:Peub4} is an achievable error exponent for the error probability caused by a particular $A$ being detected in favor of $A^{\ast}$ and the terms $k_n H_2(\gamma) + \gamma \rho k_n v(n)$ correspond to the cardinality of all possible $A$ leading to the error event $\Ecal_{\gamma}$.

By~\eqref{eq:m_lambda_rho}, it is straightforward to show that
\begin{align}
  E_0(\gamma, \rho) &= - \log m_{\lambda, \rho} (w_1, w_2)|_{w_1 = w_2 = \gamma k_n, \lambda = \frac{1}{1+\rho}}.
\end{align}
By particularizing~\eqref{eq:m_lambda} with $w_1 = w_2 = \gamma k_n$ and $\lambda = \frac{1}{1+\rho}$, we can derive $E_0(\gamma, \rho)$ explicitly as
\begin{align}
E_0(\gamma, \rho)
\label{eq:E0exp} &= \frac{\rho}{2} \log \left( 1 +\frac{\gamma k_n P' }{\rho+1 }  \right).
\end{align}
The achievable error exponent for $P(\Ecal_{\gamma})$ is determined by the minimum error exponent over the range of $\gamma$, i.e.,
\begin{align}\label{eq:Er}
E_r = \min_{ \frac{1}{k_n} \leq \gamma \leq 1 }  \max_{0 \leq \rho \leq 1} f(\gamma, \rho) .
\end{align}


\begin{lemma}\label{lemma:errexp}
Let $M$ be such that the message length $v(n ) = \log M$ is given by
\begin{align}\label{eq:msglen}
v(n) = (1- \epsilon) \frac{n}{2 k_n } \log(1 + k_n  P')  .
\end{align}
Suppose $k_n = O(n)$, there exists $n^{\ast}$ and $c_0 >0$ such that for every $n \geq n^{\ast}$,
\begin{align}
\prob\{\Ecal_{k} | A^{\ast} \} \leq e^{- c_0 n}
\end{align}
holds uniformly for all $1 \leq k \leq k_n$ and for all $|A^{\ast}|$.
\end{lemma}
\begin{IEEEproof}
See Appendix~\ref{append:pflemmaerr}.
\end{IEEEproof}

Due to Lemma~\ref{lemma:errexp}, for large enough $n$,
\begin{align}
\sum_{k=1}^{k_n} \prob\{\Ecal_{k} | A^{\ast} \} \leq k_n e^{- c_0 n} 
\end{align}
which vanishes as $n$ increases. Moreover, following the same argument as~\eqref{eq:probpow}, the second term of the RHS of~\eqref{eq:avgPe} vanishes and hence $\prob_e^{(n)}$ given by~\eqref{eq:avgPe} can be proved to vanish. As a result, Theorem~\ref{thm:logCapStage2} is established.

\subsection{Achieving the Capacity of MnAC with Random Access}
\label{subsec:achievecapacity}

In this subsection, we combine the results of Section~\ref{subsec:pfthm_logCapStage1} and Section~\ref{subsec:pfthm_logCapStage2} to prove the achievability result for Case 1 and Case 2 in Theorem~\ref{thm:logCap}. We first prove the case of unbounded $k_n$, and the case of bounded $k_n$ follows naturally. Let $\theta$ denote the limit of $\theta_n$.

\textit{Case 1) unbounded $\ell_n$ and unbounded $k_n$. }

We further divide this case into two sub-cases.

\textit{Sub-case a) $0 < \theta <1$:}
We shall show that the message length $(1 - \epsilon) B(n) $ is asymptotically achievable for every $\epsilon \in (0,1)$.

The detection errors are caused by activity identification error or message decoding error. It has been shown by~\eqref{eq:ubPB2} that with high probability the number of active users is no more than $(1+ \delta_n) k_n$. As a result, Theorem~\ref{thm:capacity_identify} and Theorem~\ref{thm:logCapStage2} conclude that the message length
\begin{align}\label{eq:msg_achieve_case0}
\frac{(1-\epsilon')(n - n_0)}{2  (1+ \delta_n)  k_n}  \log \left( 1 +  (1+ \delta_n)  k_n P \right),
\end{align}
where $n_0 = (1+\epsilon') \theta_n n$, is asymptotically achievable for any $\epsilon' >0$, when the number of active user is $(1+\delta_n) k_n$. Therefore, the message length~\eqref{eq:msg_achieve_case0} is aymptotically achievable for any fewer number of active users.

In order to prove the achievability, it suffices to show that there exists $\epsilon'$ such that the message length given by~\eqref{eq:msg_achieve_case0} is asymptotically greater than
\begin{align}\label{eq:msg_achieve_ub_case0}
(1-\epsilon) B(n) = \frac{(1 - \epsilon) ( 1 - \theta_n) n }{2 k_n} \log \left( 1 +   k_n P \right).
\end{align}
The intuition of proof is that for sufficiently large $n$, $(1+\delta_n) k_n$ is approximately $k_n$, and we can always find a small enough $\epsilon'$ such that $(1-\epsilon')(n - n_0)$ is greater than $(1-\epsilon) (1-\theta_n) n$.

We choose some small enough $\epsilon' >0$ such that
\begin{align}\label{eq:eps1_case0}
(1-\epsilon')^2  - \epsilon' (1-\epsilon')^2 \frac{ 1 + \theta }{1 - \theta} > 1 - \epsilon.
\end{align}
This is feasible because the left-hand side of~\eqref{eq:eps1_case0} is equal to~1 if $\epsilon'=0$.

Since $\log \left( 1 +  (1+ \delta_n)  k_n P \right) /\log(1+ k_n P) \to 1$ and $\delta_n \to 0$ as $n$ increases, we have
\begin{align}\label{eq:asymp_log}
\frac{\log \left( 1 +  (1+ \delta_n)  k_n P \right) }{(1+\delta_n)} \geq_n (1-\epsilon')\log(1+ k_n P).
\end{align}
The difference between~\eqref{eq:msg_achieve_case0} and $(1-\epsilon ) B(n)$ is equal to
\begin{align}
\nonumber &\frac{(1-\epsilon')(n - n_0)}{2  (1+ \delta_n)  k_n}  \log \left( 1 +  (1+ \delta_n)  k_n P \right) - (1-\epsilon) B(n) \\
& \geq_n \left[ \frac{(1-\epsilon')^2 (1 - n_0 /n)}{1 - \theta_n} - (1-\epsilon)  \right] B(n) \\
&  = \left[ (1-\epsilon')^2  - \epsilon' (1-\epsilon')^2 \frac{\theta_n}{1 - \theta_n} - (1-\epsilon) \right] B(n) \\
\label{eq:lb_theta_n} & \geq_n \left[ (1-\epsilon')^2  - \epsilon' (1-\epsilon')^2 \frac{ 1 + \theta }{1 - \theta} - (1-\epsilon) \right] B(n)
\end{align}
where~\eqref{eq:lb_theta_n} is due to $\theta_n \leq_n (1+\theta)/2$.  By~\eqref{eq:eps1_case0}, the RHS of~\eqref{eq:lb_theta_n} is greater than zero. It means that for large enough $n$, the achievabile message length~\eqref{eq:msg_achieve_case0} is greater than $(1-\epsilon) B(n)$, which establishes the achievability.

\textit{Sub-case b) $\theta = 0$:}
The proof for the case of vanishing $\theta_n$ is analogous. We shall show that message length $(1-\epsilon) B_1 (n)$ is asymptotically achievable for all $\epsilon \in (0,1)$.

The number of active users is no more than $(1+ \delta_n) k_n$ with high probability. As a result, Theorem~\ref{thm:capacity_identify} and Theorem~\ref{thm:logCapStage2} conclude that the message length
\begin{align}\label{eq:msg_achieve_case1}
\frac{(1-\epsilon')(n - n_0)}{2  (1+ \delta_n)  k_n}  \log \left( 1 +  (1+ \delta_n)  k_n P \right),
\end{align}
where $n_0 = \epsilon' n$, is asymptotically achievable for all $\epsilon' >0$.

In order to prove Theorem~\ref{thm:logCap}, it suffices to show that there exists $\epsilon'$ such that the message length given by~\eqref{eq:msg_achieve_case1} is asymptotically greater than
\begin{align}
(1 - \epsilon) B_1(n)  = (1-\epsilon)   \frac{ n }{2 k_n} \log \left( 1 +   k_n P \right).
\end{align}

Choose some small enough $\epsilon' >0$ such that
\begin{align}\label{eq:eps1_case1}
(1-\epsilon')^3  > (1-\epsilon).
\end{align}
The difference between~\eqref{eq:msg_achieve_case1} and $(1-\epsilon ) B_1(n)$ is equal to
\begin{align}
\nonumber &\frac{(1-\epsilon')(n - n_0)}{2  (1+ \delta_n)  k_n}  \log \left( 1 +  (1+ \delta_n)  k_n P \right) - (1-\epsilon) B_1(n) \\
\label{eq:lb_diff_subcase1} & \geq_n \left[ (1-\epsilon')^2 (1 - n_0 /n) - (1-\epsilon)  \right] B_1(n) \\
\label{eq:lb_diff_subcase1_1} &  = \left[ (1-\epsilon')^3  - (1-\epsilon) \right] B(n) ,
\end{align}
where~\eqref{eq:lb_diff_subcase1} is due to~\eqref{eq:asymp_log}. By the choice of $\epsilon'$ given by~\eqref{eq:eps1_case1},~\eqref{eq:lb_diff_subcase1_1} is greater than zero. It concludes that for large enough $n$, the achievable message length~\eqref{eq:msg_achieve_case1} is greater than $(1-\epsilon) B_1(n)$, which establishes the achievability.

\textit{Case 2) unbounded $\ell_n$ and bounded $k_n$.}

In this case, there is nonvanishing probability that the number of active users is equal to any finite number. The number of active users is no longer fewer than $(1+\delta_n) k_n$ with high probability. Let $s_n$ be any increasing sequence. There is high probability that the number of users is fewer than $(1+\delta_n) s_n$. As a result, by treating $s_n$ as the unbounded $k_n$ as in Case 1, we can apply the established achievable results for Case 1. The achievability result for Case 2 is summarized in the following theorem.

\begin{theorem}\label{thm:achiev_bounded_k}
Suppose $\ell_n$ is unbounded and $k_n$ is bounded. Let $s_n$ be any increasing sequence satisfying $s_n = O(n)$, $\ell_n e^{- \delta s_n} \to 0$ for every $\delta >0$ and
	\begin{align}
	\lim_{n \to \infty} \frac{ 2 \ell_n H_2 (s_n/\ell_n)  }{n \log(1 + s_n  P) } <1.
	\end{align}
	Then every message length given by
	\begin{align}\label{eq:msg_len_bounded_k}
	   (1 - \epsilon) \left(\frac{ n}{2 s_n}
     \log(1 + s_n P) - H_2 \left(\frac{s_n}{\ell_n}\right) \right)
	\end{align}
	is asymptotically achievable.
\end{theorem}
\begin{IEEEproof}
	See Appendix~\ref{sec:proof_bounded_k}.
\end{IEEEproof}

Recall that according to Theorem~\ref{thm:capacityKMac}, any message length growing linearly in $n$ is not achievabile under the conditions of Theorem~\ref{thm:achiev_bounded_k}. Theorem~\ref{thm:achiev_bounded_k} states that the message length given by~\eqref{eq:msg_len_bounded_k}, which grows sublinearly in $n$, is achievable.

\section{Successive Decoding for MnAC}
\label{sec:succ_decode}

In conventional multiaccess channels, the sum capacity can be achieved by successive decoding. A natural question is: Can the sum capacity of the MnAC be achieved using successive decoding? We consider the system model where all users have the same power constraints, assuming no random activity and the number of users being $k_n = a n$ for some $a >0$.
We provide a negative answer for the case where Gaussian random codes are used and successive decoding is applied.
Throughout the discussion in this section, we do not insist on symmetric message lengths.

Suppose Gaussian random codes are used, i.e., each user generates its codewords as i.i.d. Gaussian random variables with zero mean and variance $P$. Thus the codewords of other users look like Gaussian noise to any given user. The first user to be decoded has the largest interference from all the other $k_n -1$ users and its signal-to-interference-plus-noise ratio (SINR) is $Q = P/(1 + (k_n -1)P)$. Suppose the first user transmits with message length
\begin{align}\label{eq:msg_len_succ}
v(n) = (1- \epsilon) n C,
\end{align}
where $C = \frac{1}{2} \log (1 +Q)$. We will show that the error probability is strictly bounded from zero. The intuition is that the error probability usually decays at the rate of $\exp(- \delta n C)$, where $\delta$ is some positive constant dependent on $\epsilon$. In the MnAC setting, if the interference due to many users is so large that $n C$ converges to a finite constant, the error exponent is not large enough to drive the error probability to zero as the blocklengh increases.

\begin{lemma}\label{lemma:err_succ_equal}
Suppose Gaussian random codes are used and successive decoding is applied. There exist universal constants $d_1>0$ and $d_2>0$, such that the error probability of the first user is lower bounded as
\begin{align}
\prob_e^{(n)} \geq \mathsf{Q} (x) e^{- \frac{ d_1 T x^3}{ S^{3/2} }} \left( 1 - \frac{ d_2 T x}{S^{3/2}} \right) - e^{ - (\lambda - 1) (n-1) \epsilon C },
\end{align}
where  $\mathsf{Q} (x) = \frac{1}{\sqrt{2 \pi}} \int_x^{\infty} \exp( - \frac{u^2}{2}) du$, $S = 2 n Q (2 +Q)$,
\begin{align}
x = 2 (\lambda \epsilon n + 1 - \lambda \epsilon) C (1+Q) S^{-\frac12},
\end{align}
and
\begin{align}\label{eq:succ_T}
T = n \Exp \left\{ (-Q  (1 - Z^2) - 2 \sqrt{Q} Z)^3 \right\}
\end{align}
where $Z$ is a standard Gaussian random variable.
\end{lemma}
\begin{IEEEproof}
See Appendix~\ref{append:proof_err_succ_equal}.
\end{IEEEproof}

Let $k_n = a n$ for some constant $a >0$. Then, as $n \to \infty$,
we have
$n Q \to {1}/{a}$,
$S \to {4}/{a}$,
$T \to 0$,
$n C \to {1}/{(2a)}$, and
$x \to {\epsilon \lambda}/{(2 \sqrt{a})}$.
Therefore,
\begin{align}\label{eq:lim_err_succ_equal}
\lim_{n \to \infty } \prob_e^{(n)} \geq \mathsf{Q} \left( \frac{\epsilon \lambda}{2 \sqrt{a}} \right) - e^{ - \frac{ (\lambda -1) \epsilon}{2 a}}.
\end{align}

\begin{figure}
  \centering
  \includegraphics[width=\columnwidth]{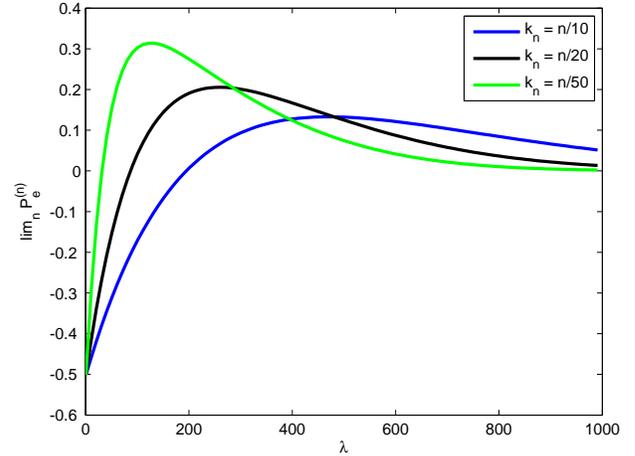}\\
  \caption{Lower bound of error probability given by~\eqref{eq:lim_err_succ_equal} for successive decoding with $\epsilon = 10^{-3}$.}\label{fig:err_equal}
\end{figure}

Using the lower bound $\mathsf{Q} (x) \geq \frac{1}{\sqrt{2 \pi}} \left( \frac{1}{x} - \frac{1}{x^3} \right) e^{- x^2 /2} $, it can be seen that when the exponential term is dominating, there exists some small enough $\lambda \epsilon$ such that the first term in~\eqref{eq:lim_err_succ_equal} is greater than the second term. In this case, the error probability is strictly bounded away from zero. Fig.~\ref{fig:err_equal} plots the numerical results of the RHS of~\eqref{eq:lim_err_succ_equal} for different values of $a$ and $\lambda$. It can be seen that for the different values of $a$, there exists some $\lambda$ that makes the lower bound of error probability~\eqref{eq:lim_err_succ_equal} strictly greater than zero.

\section{MnAC with Heterogeneous User Groups}
\label{sec:hetergain}

In this section, we will generalize the characterization of capacity region to the case where groups of users have heterogeneous channel gains and activity patterns. Suppose $\ell_n$ users can be divided into a finite number of $J$ groups, where group $j$ consists of $\beta^{(j)} \ell_n$ users with $\sum_{j=1}^J \beta^{(j)} = 1$. Further assume every user in group $j$ has the same power constraint $P^{(j)}$. Each user in group $j$ transmits with probability $\alpha_n^{(j)}$. We refer to such MnAC with heterogeneous channel gains and activity patterns as the configuration $\left( \{\alpha^{(j)}_n \}, \{\beta^{(j)}\}, \{P^{(j)}\}, \ell_n \right)$.
The error probability is defined as the probability that the receiver incorrectly detects the message of any user in the system. The problem is what is the maximum achievable message length for users in each group such that the average error probability vanishes.

\begin{definition}[Asymptotically achievable message length tuple]
Consider a MnAC of configuration $\left( \{\alpha^{(j)}_n \}, \{\beta^{(j)}\}, \{P^{(j)}\}, \ell_n \right)$. A sequence of $\big( \lceil \exp(v^{(1)} (n)) \rceil$, $\cdots$, $\lceil \exp(v^{(J)} (n)) \rceil, n \big) $ code  for this configuration consists of a  $\left(  \lceil \exp(v^{(j)} (n)) \rceil,n \right)$ symmetry code for every user in group $j$ according to Definition~\ref{def:MNCode}, $j = 1, \cdots, J$. We say a message length tuple $\left( v^{(1)} (n), \cdots, v^{(J)} (n) \right)$ is asymptotically achievable if there exists a sequence of $\left( \lceil \exp(v^{(1)} (n)) \rceil, \cdots, \lceil \exp(v^{(J)} (n)) \rceil, n \right) $ codes such that the average error probability vanishes as $n \to \infty$.
\end{definition}

\begin{definition}[Capacity region of the many-access channel]
Consider a MnAC of configuration $\left( \{\alpha^{(j)}_n \}, \{\beta^{(j)}\}, \{P^{(j)}\}, \ell_n \right)$. The capacity region is the set of asymptotically achievable message length tuples. In particular, for every $\left(B^{(1)}(n), \cdots, B^{(J)}(n) \right)$ in the capacity region, if the users transmit with message length tuple $\left((1-\epsilon)B^{(1)}(n), \cdots, (1-\epsilon) B^{(J)}(n) \right)$, the average error probability vanishes as $n \to \infty$. If users transmit according to a message-length tuple outside the capacity region, then the communication cannot be reliable.
\end{definition}

\begin{theorem}\label{thm:caphetgain}
Consider a MnAC of configuration $\left( \{\alpha^{(j)}_n \}, \{\beta^{(j)}\}, \{P^{(j)}\}, \ell_n \right)$. Suppose $\ell_n \to \infty$ and for every $j\in\{1,\dots,J\}$, $\alpha_n^{(j)} \to \alpha^{(j)} \in [0,1]$. Let the average number of active users in group $j$ be $k_n^{(j)} = \alpha_n^{(j)} \beta^{(j)}  \ell_n = O(n) $, such that $\ell_n e^{- \delta k_n^{(j)}} \to 0$ for all $\delta >0$ and $j = 1, \cdots, J$. Let $\theta_n^{(j)}$ be defined as
\begin{align}\label{eq:thetanj}
  \theta_n^{(j)} = \frac{2 \beta^{(j)} \ell_n H_2 \left( \alpha_n^{(j)} \right)  }{   n \log   k_n^{(j)} }
\end{align}
and let $\theta^{(j)}$ denote its limit. Suppose $\log k_n^{(j_1)} / \log k_n^{(j_2)} \to 1$ for any $j_1, j_2 \in \{1, \cdots, J\}$. If $\sum_{j=1}^J \theta^{(j)} <1$, then the message length capacity region is characterized as
\begin{align}\label{eq:caphetgain}
\sum_{j=1}^J k_n^{(j)} B^{(j)} (n) \leq  \frac{n}{2} \log \left(  \sum_{j=1}^J k_n^{(j)} \right) - \sum_{j=1}^J \beta^{(j)} \ell_n H_2 \left( \alpha_n^{(j)} \right).
\end{align}
If $\sum_{j=1}^J \theta^{(j)} >1$, then some user cannot transmit a single bit reliably.
\end{theorem}

As far as the asymptotic message lengths are concerned, the impact of the transmit power is inconsequential. Also, the only limitation on the message is their weighted average. This is in contrast to the classical multiaccess channel, where the sum rate of each subset of users is subject to a separate upper bound in general.

\subsection{Converse}

The proof of converse follows similarly as in Section~\ref{sec:converse}. We only sketch the proof here. Consider the system model described by~\eqref{eq:systemmodel2}. Suppose the message length transmitted by each user in group $j$ is $v^{(j)}(n)$, $j = 1, \cdots, J$. Let $\tilde{\bX}_j$ denote a vector, which stacks the vectors $\bX_k$, for all $k$ belonging to group $j$. Since there are a total of $\beta^{(j)} \ell_n$ users in group $j$ and the distributions of $\bX_k$ are the same for all $k$ in the same group $j$, we have
\begin{align}
 H \left( \tilde{\bX}_j \right) &= \beta^{(j)} \ell_n H (\bX_k) \\
\label{eq:H_j_group_3} &= \beta^{(j)} \ell_n \left(H_2 \left(\alpha_n^{(j)} \right) + \alpha_n^{(j)} v^{(j)}(n) \right).
\end{align}
Let $\grp$ denote an arbitrary subset of $\{1, \cdots, J\}$ and let $\overline{\grp}$ denote $\{1,\dots,J\}\setminus\grp$.
Further denote $\tilde{\bX}_{ \grp}$ as the vector consisting of $\{\tilde{\bX}_j : j \in \grp \}$. Thus,
\begin{align}
\label{eq:H_j_group_0} H \left( \tilde{\bX}_{ \grp} \right) & = \sum_{j \in \grp} H \left( \tilde{\bX}_j \right).
\end{align}
Applying the chain rule, we have
\begin{align}
H &\left( \tilde{\bX}_{\grp} \right) = I \left( \tilde{\bX}_{\grp}; \bY \right) + H \left( \tilde{\bX}_{\grp} | \bY \right) \\
&= H \left( \tilde{\bX}_{\grp} | \tilde{\bX}_{\overline{\grp}} \right) -  H \left( \tilde{\bX}_{\grp} | \bY \right)  + H \left( \tilde{\bX}_{\grp} | \bY \right) \\
\label{eq:H_j_group_1} & \leq I \left( \tilde{\bX}_{\grp}; \bY | \tilde{\bX}_{\overline{\grp}} \right) + H \left( \tilde{\bX}_{\grp} | \bY \right).
\end{align}
Following the argument in Lemma~\ref{lemma:hy}, we have
\begin{align}
\label{eq:I_j_group} I \left( \tilde{\bX}_{\grp}; \bY | \tilde{\bX}_{\overline{\grp}} \right) \leq \frac{n}{2} \log \left( 1 + \sum_{j \in \grp}  k_n^{(j)} P^{(j)} \right).
\end{align}
In order to achieve vanishing error probability, following the argument in Lemma~\ref{lemma:HBE}, we have
\begin{align}
\label{eq:H_j_group_2} H \left( \tilde{\bX}_{\grp} | \bY \right) = o \left( \sum_{j \in \grp}  k_n^{(j)}  v^{(j)}(n) + \beta^{(j)} \ell_n H_2 \left( \alpha^{(j)}_n \right) \right).
\end{align}
Combining~\eqref{eq:H_j_group_3},~\eqref{eq:H_j_group_0},~\eqref{eq:H_j_group_1},~\eqref{eq:I_j_group}, and~\eqref{eq:H_j_group_2}, we have
\begin{align}  \label{eq:ub_msg_tuple}
  \begin{split}
    (1-\epsilon)
    & \sum_{j \in \grp } \left[  k_n^{(j)} v^{(j)} (n) + \beta^{(j)} \ell_n H_2 \left( \alpha_n^{(j)} \right) \right] \\
    & \le \frac{n}{2} \log \left( 1 +  \sum_{j \in \grp } k_n^{(j)} P^{(j)} \right) ,
  \end{split}
\end{align}
for all $\epsilon >0$ and large enough $n$.

Since the power in each group is bounded, we have
\begin{align}
  \lim_{n\to\infty} \frac
  { \log \left( 1 + \sum_{j \in \grp } k_n^{(j)} P^{(j)} \right) }
  { \log \sum_{j \in \grp } k_n^{(j)} }
  = 1 .
\end{align}
Thus,~\eqref{eq:ub_msg_tuple} implies that for every $\epsilon >0$,
\begin{align} \label{eq:converse_j_group_1}
  \begin{split}
    &\sum_{j \in \grp }  k_n^{(j)} v^{(j)} (n) \leq \\
    &\; (1+ \epsilon) \frac{n}{2 } \log \left( \sum_{j \in \grp } k_n^{(j)}  \right) -  \sum_{j \in \grp } \beta^{(j)} \ell_n H_2 \left( \alpha_n^{(j)} \right).
  \end{split}
\end{align}
As in~\eqref{eq:symmcapacity_2}, we have dropped the power terms in the capacity expression to ease the rest of the proof. By~\eqref{eq:converse_j_group_1}, we have
\begin{align} \label{eq:converse_j_group_2_0}
    \sum_{j \in \grp }
    k_n^{(j)} v^{(j)} (n) \leq
    \bigg( 1+ \epsilon - \sum_{j \in \grp } \theta_n^{(j)} \xi_n^{(\grp,j)}  \bigg)
    \frac{n}{2 } \log \sum_{j \in \grp } k_n^{(j)} ,
\end{align}
where
\begin{align}
\xi_n^{(\grp, j)}  = \frac{ \log k_n^{(j)} }{ \log \sum_{j \in \grp } k_n^{(j)} }.
\end{align}
Given any $\grp_1, \grp_2 \subseteq \{1, \cdots, J\}$, we have
\begin{align}
  & \frac{ \log \left(  \min_{j \in \grp_1 } k_n^{(j)}  \right)}{  \log \left(  \max_{j \in \grp_2 } k_n^{(j)} \right) + \log J }
  \leq
  \frac{ \log\sum_{j\in\grp_1}k_n^{(j)} }{ \log\sum_{j\in\grp_2}k_n^{(j)} } \\
  \label{eq:log_ratio}
  & \qquad\qquad\qquad \leq
\frac{ \log \left(   \max_{j \in \grp_1 } k_n^{(j)} \right)  + \log J }{  \log \left(  \min_{j \in \grp_2 } k_n^{(j)} \right)}.
\end{align}
Taking the limit of $n \to \infty$ on both sides of~\eqref{eq:log_ratio}, by the assumption that $\log k_n^{(j_1)} / \log k_n^{(j_2)} \to 1$, $\forall j_1, j_2$, we have
\begin{align}\label{eq:log_ratio_1}
  \frac
  { \log \sum_{j \in \grp_1 } k_n^{(j)} }
  { \log \sum_{j \in \grp_2 } k_n^{(j)} }
  \to 1.
\end{align}
It implies that
\begin{align} \label{eq:xin}
  \lim_{n\to\infty} \xi_n^{(\grp,j)} = 1, \quad \forall j \in \grp.
\end{align}
If $\sum_{j=1}^J \theta^{(j)} >1$, particularizing~\eqref{eq:converse_j_group_2_0} with $\grp = \{1, \cdots, J \}$ implies that for large enough $n$, $v^{(j)} (n) = 0$ for all $j = 1, \cdots, J$.

If $\sum_{j=1}^J \theta^{(j)} <1$, the achievable message length can be further upper bounded as
\begin{align}
\label{eq:converse_j_group_2}\sum_{j \in \grp } k_n^{(j)} v^{(j)} (n) &\leq
  \left( 1 + \frac{\epsilon}{1-  \sum_{j \in \grp } \theta_n^{(j)} \xi_n^{(\grp,j)}  }  \right) B_{\grp} (n),
\end{align}
where
\begin{align}\label{eq:Bj}
B_{\grp} (n) =  \frac{n}{2 } \log \left(  \sum_{j \in \grp } k_n^{(j)}  \right) -  \sum_{j \in \grp } \beta^{(j)} \ell_n H_2 \left( \alpha_n^{(j)} \right) .
\end{align}
Applying~\eqref{eq:converse_j_group_2} with $\grp = \{ 1, \cdots, J \}$ and $\xi_n^{(\grp,j)} \to 1$, the achievable message length tuple must satisfy
\begin{align}
\sum_{j \in \{1, \cdots, J \} } k_n^{(j)} v^{(j)} (n) \leq (1+\epsilon) B_{ \{1, \cdots, J \} } (n)
\end{align}
for all $\epsilon >0$. Thus, the converse part of Theorem~\ref{thm:caphetgain} is established.

By~\eqref{eq:converse_j_group_2}, any achievable message length tuple must satisfy
\begin{align}
\sum_{j \in \grp } k_n^{(j)} v^{(j)} (n) \leq
(1+\epsilon) B_{\grp} (n)
\end{align}
for all $\grp \subseteq \{1, \cdots, J\}$. However, in the regime of unbounded $k_n$,~\eqref{eq:converse_j_group_2} implies that these constraints are dominated by the one for $\grp = \{1, \cdots, J\}$, because $B_{\grp} (n) \geq_n B_{\{1, \cdots, J\}} (n)$ for all $\grp \subseteq \{1, \cdots, J\}$.

\subsection{Achievability}

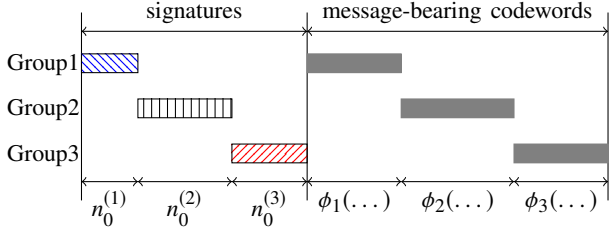
\begin{figure}
  \centering
  \begin{tikzpicture}
    \small
    \draw (1,-.25) -- (1,2.25);
    \draw (4,-.25) -- (4,2.25);
    \draw (8,-.25) -- (8,2.25);
    \node[anchor=south west] at (-0.1,1.3) {Group1};
    \node[anchor=south west] at (-0.1,.7) {Group2};
    \node[anchor=south west] at (-0.1,0.1) {Group3};
    \draw[<->] (1,0) -- (1.75,0) node[pos=0.5,below]{$n_0^{(1)}$};
    \draw[<->] (1.75,0) -- (3,0) node[pos=0.5,below]{$n_0^{(2)}$};
    \draw[<->] (3,0) -- (4,0) node[pos=0.5,below]{$n_0^{(3)}$};
    \draw[<->] (4,0) -- (5.25,0) node[pos=0.5,below]{$\phi_1(\dots)$}; 
    \draw[<->] (5.25,0) -- (6.75,0) node[pos=0.5,below]{$\phi_2(\dots)$}; 
    \draw[<->] (6.75,0) -- (8,0) node[pos=0.5,below]{$\phi_3(\dots)$}; 
    \draw[<->] (1,2) -- (4,2) node[midway,above]{signatures};
    \draw[<->] (4,2) -- (8,2) node[midway,above]{message-bearing codewords};
    \draw[pattern=north west lines,pattern color=blue] (1,1.45) rectangle (1.75,1.7);
    \draw[pattern=vertical lines,pattern color=black] (1.75,.85) rectangle (3,1.1);
    \draw[pattern=north east lines,pattern color=red] (3,0.25) rectangle (4,0.5);
    \filldraw[color=gray] (4,1.45) rectangle (5.25,1.7);
    \filldraw[color=gray] (5.25,.85) rectangle (6.75,1.1);
    \filldraw[color=gray] (6.75,0.25) rectangle (8,0.5);
  \end{tikzpicture}
	\caption{Transmission scheme for $J=3$ groups.}\label{fig:transmit_j_group}
\end{figure}

We need to prove that the region of the achievable message length tuple covers the region specified by~\eqref{eq:caphetgain}. In particular, we will show that the message length tuple satisfying
\begin{align}
\nonumber & \sum_{j=1}^J k_n^{(j)} v^{(j)} (n) \leq  \\
\label{eq:caphetgain_2}&(1-\epsilon) \left[ \frac{n}{2} \log \left(   \sum_{j=1}^J k_n^{(j)} \right) - \sum_{j=1}^J \beta^{(j)} \ell_n H_2 \left( \alpha_n^{(j)} \right) \right]
\end{align}
is asymptotically achievable for all $\epsilon >0$.

One achievable scheme is to detect active users in each group and their transmitted messages in a time-division manner. In particular, in the first stage, we let users in group~1 transmit the signatures before group~2, and so on. The signature length transmitted by users in group $j$ is $n_0^{(j)}$. 
In the second stage, we let each group share the remaining time resource $n-\sum_{j=1}^{J} n_0^{(j)}$. Users in group~1 transmit their message-bearing codewords before group~2, and so on. The time resource allocated to group $j$ in the second stage is $\phi_j \left( n-\sum_{j=1}^{J} n_0^{(j)} \right)$, where $\phi_j \geq 0$ and $\sum_{j=1}^J \phi_j = 1$.
According to the group order, the receiver first identifies active users and then decodes the transmitted messages.  The overall scheme is illustrated in Fig.~\ref{fig:transmit_j_group}.

Let $\theta_n^{(j)}$ be given by~\eqref{eq:thetanj}, which can be regarded as the fraction of channel uses dedicated to the identification of active users in group $j$.
According to Theorems~\ref{thm:capacity_identify} and~\ref{thm:logCapStage2}, the message length tuple satisfying
\begin{align}\label{eq:achiev_msg_tuple}
v^{(j)}(n) = (1-\epsilon') \phi^{(j)} \frac{n-\sum_{j' =1}^{J} n_0^{(j')}}{2 k_n^{(j)}} \log k_n^{(j)} ,
\end{align}
where
\begin{align}\label{eq:NTj}
  n_0^{(j)} =
  \begin{cases}
    n (1+ \epsilon'/2) \theta_n^{(j)}, \;\; & \text{ if } \theta^{(j)} >0 \\
    n \epsilon' / (2 J), & \text{ if } \theta^{(j)} =0
  \end{cases}
\end{align}
is achievable for all $\epsilon' \in (0,1)$.

If $\theta^{(j')} >0$, by~\eqref{eq:log_ratio_1},
\begin{align}
  \frac{n_0^{(j')}}{2} \log k_n^{(j)}
  & = \left(1+\frac{\epsilon'}{2}\right)
  \beta^{(j')} \ell_n H_2 \left( \alpha_n^{(j')} \right) \frac{\log k_n^{(j)}} {\log k_n^{(j')}} \\
  &\leq_n  (1+\epsilon') \beta^{(j')} \ell_n H_2 \left( \alpha_n^{(j')} \right) .
\end{align}
If $\theta^{(j')} =0$,
\begin{align}
  \frac{n_0^{(j')}}{2} \log k_n^{(j)} = \frac{\epsilon' }{2 J} \frac{n}{2} \log k_n^{(j)} .
\end{align}
Therefore,
\begin{align}
  \begin{split}
    \sum_{j'=1}^J & \frac{n_0^{(j')}}{2} \log ( k_n^{(j)} )  \leq_n  \\
    & \frac{\epsilon'}{2}  \frac{n}{2} \log k_n^{(j)} + \sum_{j'=1}^J (1+\epsilon') \beta^{(j')} \ell_n H_2 \left( \alpha_n^{(j')} \right) .
  \end{split}
\end{align}
By~\eqref{eq:xin}, the achievable message length described by~\eqref{eq:achiev_msg_tuple} satisfies
\begin{align}
  \nonumber & k_n^{(j)} v^{(j)}(n)  \\
  \nonumber &\geq_n   (1-\epsilon')  \phi^{(j)} \times
  \Bigg[  \left(1-\frac{\epsilon'}2\right)  \frac{n}{2 } \log k_n^{(j)} \\
  &\qquad - (1+\epsilon')   \sum_{j'=1}^J \beta^{(j')} \ell_n H_2 \left( \alpha_n^{(j')} \right)  \Bigg] \\
  &\geq_n \phi^{(j)} (1-\epsilon) \nonumber\\
  &\quad \times \left[ \frac{n}{2} \log \left(   \sum_{j'=1}^J k_n^{(j')} \right) - \sum_{j'=1}^J \beta^{(j')} \ell_n H_2 \left( \alpha_n^{(j')} \right) \right]
  \label{eq:lb_betav}
\end{align}
for some small enough $\epsilon'$ and all $j = 1, \cdots, J$.

Since~\eqref{eq:lb_betav} holds for any $\phi^{(j)}>0$, $j=1,\dots,J$,
the region spanned by the achievable message tuple~\eqref{eq:achiev_msg_tuple} covers the region specified by~\eqref{eq:caphetgain_2}. The achievability result is thus established.

\section{Conclusion}
\label{sec:conclude}

In this paper, we have proposed a model of many-access channel, where the number of users scales with the coding blocklength as a first step towards the study of many-user information theory. New notions of achievable message length and symmetric capacity have been defined. The symmetric capacity of a Gaussian many-access channel is described as a function in the channel uses, consisting of two terms. The first term is the symmetric capacity of many-access channel with knowledge of the set of active users and the second term can be regarded as the cost of user identification in random access channels. Separate identification and decoding has been shown to be capacity achieving.\footnote{This does not apply in general to non-Gaussian channels, e.g., the OR many-access channel~\cite{zhang2017or}.} The detection scheme can be extended to achieve the capacity region of a many-access channel with a finite number of groups experiencing different channel gains.

The results presented in this paper reveal the capacity growth in the asymptotic regime. A many-user information theory for finite but large number of users and finite but large block length remains to be developed, the challenge of which is hard to overestimate (see, e.g.,~\cite{polyanskiy2010channel_coding, molavianjazi2014second}).

With the advent of the Internet of Things, 
where a large population of users wish to communicate over a shared medium, there has been renewed interests to design uncoordinated multiple access protocols. Previous works have focused on efficiently identifying randomly activated users~\cite{stefanovic2016identifying} and maximizing the system throughput~\cite{paolini2015coded,madala2015uncoordinated,taghavi2016design,xie2015many,chen2017sparse}. The many-access channel model here has provided the fundamental limits of the channel capacity. The capacity result and the compressed sensing based identification technique provide guidance for the design of optimal coding and signal processing algorithms.

\appendices

\section{Proof of Lemma~\ref{lemma:hy}}
\label{append:Lemmahy}

To upper bound the input-output mutual information of the white Gaussian noise channel, it sufficies to identify the power constraint on the input signal $\bss \bX$ based on the power constraint~\eqref{eq:powerconst} on $\bss$ and the structure of the binary vector $\bX$.

According to the distribution of $\bX$, we can obtain the marginal distribution of $X_i$, $i = 1, \cdots, M \ell_n$, as
$\prob \{X_{i} = 0\} = 1 - \frac{\alpha_n}{M}$ and $\prob \{ X_{i} = 1 \} = \frac{\alpha_n}{M}$. Therefore, $\Exp  \{X_{i}\} = \frac{\alpha_n}{M}$ and
\begin{align}
  \Exp  \{ X_{i} X_j \} =
\begin{cases}
  \frac{\alpha_n}{M} & \text{if } i =j \\
  0 & \text{if } \exists l, \text{ s.t. } i,j \in I(\ell), \, i\ne j \\
  \left( \frac{\alpha_n}{M} \right)^2 & \text{otherwise}
\end{cases}
\end{align}
where we let the indices corresponding to transmitter $\ell$ be $I(\ell) =\{(\ell-1) M +1, \cdots,  \ell M \}$, $\ell = 1, \cdots, \ell_n$. Thus, the covariance matrix $\bK = \Exp  \left\{ (\bX - \Exp  \bX) (\bX - \Exp  \bX)^T \right\}$ can be calculated as
\begin{align}
  \bK_{ij} =
  \begin{cases}
    \frac{\alpha_n}{M} \left( 1 - \frac{\alpha_n}{M} \right) & \text{if } i =j \\
    - \left(\frac{\alpha_n}{M} \right)^2
    &  \text{if } \exists l, \text{ s.t. } i,j \in I(\ell), \, i\ne j \\
    0 & \text{otherwise}.
\end{cases}
\end{align}

Let $tr(\cdot)$ find the trace of a matrix. The power constraint on the codewords induces the power constraint on $\bss \bX$ as
\begin{align}
\nonumber & tr \left( \bss \bK \bss^T \right) \\
 & = tr \left( \bK \bss^T  \bss \right)  \\
 & = \sum_{i=1}^{M \ell_n} \sum_{j=1}^{M \ell_n} \sum_{k=1}^n K_{ij} s_{ki} s_{kj} \\
\nonumber &= \sum_{k=1}^n \left[  \frac{\alpha_n}{M} \left( 1 - \frac{\alpha_n}{M} \right) \sum_{i=1}^{M \ell_n} s_{ki}^2 - \right. \\
& \qquad \qquad \left. \left( \frac{\alpha_n}{M} \right)^2 \sum_{\ell=1}^{\ell_n} \sum_{i,j \in I(\ell):i\ne j} s_{ki} s_{kj} \right] \\
& =  \sum_{k=1}^n \left[  \frac{\alpha_n}{M} \sum_{i=1}^{M \ell_n} s_{ki}^2 - \left( \frac{\alpha_n}{M} \right)^2 \sum_{\ell=1}^{\ell_n}
\left(\sum_{i\in I(\ell)} s_{ki} \right)^2
 \right] \\
\label{eq:sum_square} & \leq \frac{n \alpha_n}{M}  \sum_{i=1}^{M \ell_n} \frac{1}{n} \sum_{k=1}^n s_{ki}^2 \\
& \leq k_n n P,
  \label{eq:tr<=}
\end{align}
where~\eqref{eq:tr<=} is due to the power constraint $ \frac{1}{n} \sum_{k=1}^n s_{ki}^2 \leq P$.

Since $\bX \to \bss \bX \to \bY$ forms a Markov chain, we can obtain an upper bound of $I(\bX; \bY)$ as
\begin{align}
\label{eq:ubIXYStep1}I(\bX; \bY) &\leq I (\bss \bX; \bY) \\
& \leq \max_{ tr \left( \bss \bK \bss^T \right)   \leq k_n n P } I (\bss \bX; \bY) \\
\label{eq:maxI} & \leq \frac{n}{2} \log(1 + k_n P)  ,
\end{align}
where~\eqref{eq:maxI} follows by the results on parallel Gaussian channels~\cite[Chapter 10]{cover2006elements}.


\section{Proof of Lemma~\ref{lemma:HBE}}
\label{append:lemmahbe}

Since $H \left(\bX | E = 0, \bY, 1 \left\{ \bX \in \Bcal_M^{\ell_n} (\delta, k_n) \right\} \right) = 0$, we can obtain
\begin{align} \label{eq:hbeStep1}
  \begin{split}
    H \Big( \bX | E, &\bY, 1 \left\{ \bX \in \Bcal_M^{\ell_n} (\delta, k_n)  \right\} \Big)  \\
    &= H(\bX | E = 1, \bY, \bX \notin \Bcal_M^{\ell_n} (\delta, k_n) ) \\
    &\qquad\qquad \times \prob\{E = 1, \bX \notin \Bcal_M^{\ell_n} (\delta, k_n) \}\\
    &+ H(\bX | E = 1, \bY, \bX \in \Bcal_M^{\ell_n} (\delta, k_n)  ) \\
    &\qquad\qquad \times \prob\{E = 1, \bX \in \Bcal_M^{\ell_n} (\delta, k_n)  \}.
  \end{split}
\end{align}
We upper bound the first term in the RHS of~\eqref{eq:hbeStep1} as follows: $\bX$ can take at most $(M+1)^{\ell_n}$ values and $\|\bX\|_0$ follows the binomial distribution ${\rm Bin} (\ell_n, \alpha_n)$ with mean $\ell_n \alpha_n = k_n$, then $  \prob \{\bX \notin \Bcal_M^{\ell_n} (\delta, k_n)  \}$ can be upper bounded by $ e^{- c(\delta) k_n} $~\cite{arratia1989tutorial}, where $c(\delta)$ is some constant depending on $\delta$ by the large deviations for binomial distribution. Then
\begin{align}
\nonumber & H(\bX | E\!=\!1, \bY, \bX \notin \Bcal_M^{\ell_n} (\delta, k_n)  )  \prob\{E\!=\!1, \bX \notin \Bcal_M^{\ell_n} (\delta, k_n) \} \\
& \qquad \leq  e^{- c(\delta) k_n} \ell_n \log (M+1) \\
\label{eq:HbXStpe0}& \qquad \leq_n \log M.
\end{align}

For the second term in the RHS of~\eqref{eq:hbeStep1}, $\prob\{E = 1, \bX \in \Bcal_M^{\ell_n} (\delta, k_n)  \} \leq \prob_e^{(n)}$ and
\begin{align} \label{eq:HXE1}
  H(\bX | E = 1, \bY, \bX \in \Bcal_M^{\ell_n} (\delta, k_n)  )  \leq \log | \Bcal_M^{\ell_n} (\delta, k_n) | .
\end{align}
The cardinality of $\Bcal_M^{\ell_n} (\delta, k_n)$ is
\begin{align}
 |\Bcal_M^{\ell_n} (\delta, k_n) | &= \sum_{ j=1}^{(1+\delta) k_n} \binom{\ell_n}{j} M^j \\
  \leq\, & (1+\delta) k_n M^{(1+\delta) k_n} \max_{1 \leq j \leq (1+\delta) k_n} \binom{\ell_n}{j}.
\end{align}
If $(1+\delta) k_n \geq \frac{\ell_n}{2}$, then
\begin{align}
\max_{ 1 \leq j \leq (1+\delta) k_n} \binom{\ell_n}{j} &\leq 2^{\ell_n} \\
&\leq \exp(2 (1+\delta) k_n \log 2).
\end{align}
If $(1+\delta) k_n < \frac{\ell_n}{2}$, then
\begin{align}
\max_{ 1 \leq j \leq (1+\delta) k_n} \binom{\ell_n}{j} &\leq \binom{\ell_n}{(1+\delta) k_n} \\
&\leq \exp(\ell_n H_2((1+\delta) \alpha_n)).
\end{align}
We further upper bound $H_2((1+\delta) \alpha_n)$ in terms of $H_2(\alpha_n)$. By the mean value theorem, there exists some $\gamma'_n$ in between $\alpha_n$ and $(1+\delta)\alpha_n$ such that
\begin{align}
H_2((1+\delta) \alpha_n) -H_2(\alpha_n) = \delta \alpha_n \log \frac{1 - \gamma'_n}{\gamma'_n},
\end{align}
where $\log \frac{1-x}{x}$ is the first order derivative of $H_2 (x)$. Since $\log \frac{1-x}{x}$ is decreasing in $x$, we have
\begin{align}
  H_2((1+\delta) \alpha_n) -H_2(\alpha_n)
  &\leq \delta \alpha_n \log \frac{1 - \alpha_n}{\alpha_n} \\
  &\leq \delta H_2 (\alpha_n). \label{eq:H2_ineq}
\end{align}
As a result,
\begin{align} \label{eq:logBM}
  \begin{split}
    \log | \Bcal_M^{\ell_n} (\delta, k_n) | &\leq  \log \left( (1 + \delta) k_n \right) + (1+\delta) k_n \log M \\
    + 2 & (1+\delta) k_n \log 2+ (1+\delta) \ell_n H_2( \alpha_n) .
  \end{split}
\end{align}
Because $\log \left( (1+\delta) k_n \right) \leq (1+\delta) k_n$ for large enough $n$,~\eqref{eq:HXE1} and~\eqref{eq:logBM} imply
\begin{align}
  \begin{split}
    H(\bX| E= 1, & \bX \in \Bcal_M^{\ell_n} (\delta , k_n), \bY)  \\
\label{eq:HbXStpe1}
    & \leq_n 4  (k_n \log M +  k_n + \ell_n H_2( \alpha_n) ) .
  \end{split}
\end{align}

Combining~\eqref{eq:hbeStep1},~\eqref{eq:HbXStpe0}, and~\eqref{eq:HbXStpe1} yields the lemma.

\section{Derivation of~\eqref{eq:m_lambda}}
\label{append:derivehlamdarho}

We begin with~\eqref{eq:m_lambda_rho} and write, 
\begin{align}
  \nonumber
  & m_{\lambda, \rho} ( w_1,w_2) \\
  &= \int_{\mathbb{R}}  \Exp \left\{  p^{1- \lambda \rho}_{Y | \bS_A} ( y | \bS^a_{A^{\ast}})  \left( \Exp  \left\{ p^{\lambda}_{Y | \bS_A} ( y | \bS^a_A) \Big| \bS^a_{A^{\ast}}   \right\} \right)^{\rho} \right\} d y  \\
  \label{eq:Peub2_1}
  &=  \int_{\mathbb{R}}  \Exp \left\{  p^{1- \lambda \rho}_{Y | \bS_A} ( y | \bS^a_{A^{\ast}})  \left( \Exp  \left\{ p^{\lambda}_{Y | \bS_A} ( y | \bS^a_A) \Big| \bS^a_{A^{\ast} \backslash A_1}   \right\} \right)^{\rho} \right\} d y  \\
  \nonumber
  & = \int_{\mathbb{R}} \Exp  \left\{ \Exp \left\{  p^{ 1- \lambda \rho}_{Y | \bS_A} ( y | \bS^a_{A^{\ast}})    \Big| \bS^a_{A^{\ast} \backslash A_1} \right\} \right. \\
  \label{eq:Peub2}
  & \qquad\qquad \times \left. \left(  \Exp \left\{  p^{\lambda}_{Y | \bS_A} ( y | \bS^a_{A} ) \Big| \bS^a_{A^{\ast} \backslash A_1}   \right\} \right)^{\rho} \right\} d y
\end{align}
where~\eqref{eq:Peub2_1} follows because $A \cap A^{\ast} = A^{\ast} \backslash A_1$.

Let $Z_1 = \sum_{k \in A_1} S^a_k$, $Z_2 = \sum_{k \in A_2} S^a_k$, and $Z_3 = \sum_{k \in A^{\ast} \backslash A_1} S^a_k$. Since $|A_1| = w_1$ and $|A_2| = w_2$, we have $Z_1 \sim \mathcal{N} (0, v_1)$, $Z_2 \sim \mathcal{N} (0, v_2) $, and $Z_3 \sim \mathcal{N} (0, v_3)$, where $v_1 = w_1 P'$, $v_2 = w_2 P'$, and $v_3 = ( |A^{\ast}| - w_1) P'$.
We can write
\begin{align}
\nonumber & \Exp \left\{  p^{\lambda}_{Y | \bS_A} ( y | \bS^a_{A} ) \Big| \bS^a_{A^{\ast} \backslash A_1}  \right\} \\
& = \Exp \left\{  \left( \frac{1}{\sqrt{2 \pi} } e^{- \frac{ (y -Z_3 - Z_2)^2 }{2} } \right)^{\lambda}  \bigg| Z_3 \right\}\\
 & = \int_{\mathbb{R}} \left( \frac{1}{\sqrt{2 \pi} } e^{- \frac{ (y -Z_3 - z_2)^2 }{2} } \right)^{\lambda} \frac{1}{\sqrt{2 \pi v_2} } e^{- \frac{z_2^2}{ 2 v_2} } d z_2 \\
 & = \left( \frac{1}{\sqrt{2 \pi} } \right)^{\lambda} \sqrt{ \frac{t_3}{v_2}  }  e^{ \frac{\mu_3^2}{2 t_3} } e^{-\frac{\lambda (y - Z_3)^2 }{2} } \int_{\mathbb{R}} \frac{1}{\sqrt{2 \pi t_3}} e^{- \frac{ (z_2 - \mu_3)^2 }{2 t_3} } d z_2 \\
 & = \left( \frac{1}{\sqrt{2 \pi} } \right)^{\lambda} \sqrt{ \frac{t_3}{v_2}  }  e^{ \frac{\mu_3^2}{2 t_3} } e^{-\frac{\lambda (y - Z_3)^2 }{2} } ,
\end{align}
where $\frac{1}{t_3} = \lambda + \frac{1}{v_2}$ and $\mu_3 = \lambda (y - Z_3) t_3$.
Similarly,
\begin{align}
\nonumber & \Exp \left\{  p^{ 1- \lambda \rho}_{Y | \bS_A} ( y | \bS^a_{A^{\ast}}) \Big| \bS^a_{A^{\ast} \backslash A_1}  \right\}   \\
 &= \Exp  \left\{ \left( \frac{1}{\sqrt{2 \pi}} \right)^{1 - \lambda \rho} e^{- \frac{ (1- \lambda \rho) (y - Z_3 - Z_1)^2 }{2} }  \bigg| Z_3 \right\} \\
& = \left( \frac{1}{\sqrt{2 \pi}} \right)^{1 - \lambda \rho} \int_{\mathbb{R}} e^{- \frac{ (1- \lambda \rho) (y - Z_3 - z_1)^2 }{2} } \frac{1}{\sqrt{2 \pi v_1} } e^{- \frac{z_1^2}{ 2 v_1} } d z_1 \\
\nonumber & = \left( \frac{1}{\sqrt{2 \pi}} \right)^{1 - \lambda \rho} \sqrt{\frac{t_4}{v_1} } e^{ \frac{\mu_4^2}{2 t_4} } e^{- \frac{(1-\lambda \rho) (y - Z_3)^2 }{2} } \times \\
&\qquad \qquad  \int_{\mathbb{R}} \frac{1 }{\sqrt{ 2 \pi t_4} } e^{- \frac{(z_1 - \mu_4)^2 }{2 t_4}} d z_1 \\
& =  \left( \frac{1}{\sqrt{2 \pi}} \right)^{1 - \lambda \rho} \sqrt{\frac{t_4}{v_1} } e^{ \frac{\mu_4^2}{2 t_4} } e^{- \frac{(1-\lambda \rho) (y - Z_3)^2 }{2} },
\end{align}
 where $\frac{1}{t_4} = 1-  \lambda \rho + \frac{1}{v_1}$ and $\mu_4 = (1- \lambda \rho) (y - Z_3) t_4$.
Then
\begin{align}
\nonumber & \left(  \Exp \left\{  p^{\lambda}_{Y | \bS_A} ( y | \bS^a_{A} )   \Big| \bS^a_{ A^{\ast} \backslash A_1 }   \right\} \right)^{\rho} \Exp \left\{  p^{ 1- \lambda \rho}_{Y | \bS_A} ( y | \bS^a_{A^{\ast}}) \Big|  \bS^a_{A^{\ast} \backslash A_1}  \right\} \\
&= \frac{1}{\sqrt{2 \pi}} \left(\sqrt{ \frac{t_3}{v_2}  }\right)^{\rho}  \sqrt{\frac{t_4}{v_1} }  e^{ \frac{\rho \mu_3^2}{2 t_3} + \frac{\mu_4^2}{2 t_4} -\frac{ (y - Z_3)^2 }{2} }.
\end{align}
Plugging $\mu_3$, $t_3$, $\mu_4$ and $t_4$ yields
\begin{align}
  \frac{\mu_3^2}{t_3}
  &= \frac{\lambda^2 v_2 (y - Z_3)^2 }{1 + \lambda v_2} \\
  \frac{\mu_4^2}{t_4}
  &= \frac{(1 - \lambda \rho)^2 (y - Z_3)^2 v_1 }{1 + (1 - \lambda \rho) v_1} .
\end{align}
Let
\begin{align}
  t_0 &= \frac{1}{\sqrt{2 \pi}} \left(\sqrt{ \frac{t_3}{v_2}  }\right)^{\rho}  \sqrt{\frac{t_4}{v_1} } \\
  t_5 &= \left( 1 - \frac{\rho \lambda^2 v_2}{1 + \lambda v_2} -\frac{ (1- \lambda \rho)^2 v_1}{1 + (1 - \lambda \rho) v_1} \right)^{-1}.
\end{align}
We have
\begin{align}
\nonumber & \int_{\mathbb{R}} \Exp  \left\{ \left(  \Exp \left\{  p^{\lambda}_{Y | \bS_A} ( y | \bS^a_{A} ) \Big| \bS^a_{ A^{\ast} \backslash A_1 }   \right\} \right)^{\rho}  \times \right. \\
\nonumber & \qquad \left. \Exp \left\{  p^{ 1- \lambda \rho}_{Y | \bS_A} ( y | \bS^a_{A^{\ast}}) \Big| \bS^a_{A^{\ast} \backslash A_1} \right\}  \right\} d y \\
\nonumber & = t_0 \int_{\mathbb{R}} \int_{\mathbb{R}} \frac{1}{\sqrt{2 \pi v_3} } e^{- \frac{z_3^2}{2 v_3}} \times \\
& \qquad e^{ \frac{\rho \lambda^2 v_2 (y - z_3)^2 }{ 2( 1 + \lambda v_2)} + \frac{(1 - \lambda \rho)^2 (y - z_3)^2 v_1 }{2 ( 1 + (1 - \lambda \rho) v_1 )}  -\frac{ (y - z_3)^2 }{2} } d z_3 d y \\
& = t_0 \int_{\mathbb{R}} \sqrt{ \frac{t_5}{v_3} } e^{- \frac{z_3^2}{2 v_3}} \int_{\mathbb{R}} \frac{1}{\sqrt{2 \pi t_5} } e^{- \frac{ (y - z_3)^2 }{2 t_5} } d y d z_3 \\
& = t_0 \int_{\mathbb{R}} \sqrt{ \frac{t_5}{v_3} } e^{- \frac{z_3^2}{2 v_3}} d z_3 \\
& = \left(\sqrt{ \frac{t_3}{v_2}  }\right)^{\rho}  \sqrt{\frac{t_4 t_5}{v_1} } \\
& = \left( 1+ \lambda v_2 \right)^{- \rho /2} \left( \frac{1 + \lambda v_2}{1 + \lambda (1 - \lambda \rho)v_2 + \lambda \rho (1 - \lambda \rho) v_1} \right)^{1/2}.
\end{align}
Therefore, $m_{\lambda, \rho} ( w_1,w_2)$ is given by~\eqref{eq:m_lambda}.


\section{Proof of Lemma~\ref{lemma:minhlambdarho}}
\label{append:pflemmaminhlambdarho}

We first establish the following two lemmas that will be useful in the proof.

\begin{lemma}\label{lemma:lim_H2}
Suppose~\eqref{eq:assump_identify} holds, i.e., $\lim\limits_{\ell \to \infty} \ell e^{- \delta k_{\ell}} = 0$ for every $\delta>0$, then for every constant $\bar{w} \geq 0$,
\begin{align}\label{eq:lemma_lim_H2}
\lim_{\ell \to \infty} \frac{\ell}{k_{\ell}} H_2 \left(  \frac{\bar{w}}{\ell} \right) = 0 .
\end{align}
\end{lemma}
\begin{IEEEproof}
The case of $\bar{w} = 0$ is trivial. Suppose $\bar{w}>0$. Since $\bar{w} / \ell \to 0$,
\begin{align}
 \frac{\ell}{k_{\ell}} H_2 \left(  \frac{\bar{w}}{\ell} \right) & = \frac{\ell}{k_{\ell}} \left( \frac{\bar{w}}{\ell} \log \frac{\ell}{\bar{w}} - \left( 1 - \frac{\bar{w}}{\ell} \right) \log \left( 1 - \frac{\bar{w}}{\ell} \right) \right)\\
 & \leq_{\ell}  \frac{\ell}{k_{\ell}} \left( \frac{\bar{w}}{\ell} \log \frac{\ell}{\bar{w}} + \left( 1 - \frac{\bar{w}}{\ell} \right) \frac{2 \bar{w}}{\ell} \right) \\
\label{eq:H2w} & \leq   \frac{\bar{w}}{k_{\ell}}   \left( \log \ell - \log \bar{w} + 2 \right).
\end{align}
Since $\ell e^{- \delta k_{\ell}} \to 0$ for every $\delta>0$, we have $\ell \leq_{\ell} e^{\delta k_{\ell}}$, so that $\log \ell \leq_{\ell} \delta k_{\ell} $. This implies $(\log \ell)/ k_{\ell}\to 0$, so that the RHS of~\eqref{eq:H2w} vanishes.
\end{IEEEproof}

\begin{lemma}\label{lemma:lbhlargeK}
Suppose~\eqref{eq:assump_identify} holds for all $\delta >0$. Let $A>0$, $B>0$ and $\bar{w}\geq 1$ be constants. Let  $\{a_{\ell}\}$ and $\{b_{\ell}\}$ be two sequences that satisfy $b_{\ell} \leq a_{\ell}$, $\lim\limits_{\ell \to \infty} \frac{k_{\ell}}{a_{\ell}} = a \in [0, \infty)$, and $\lim\limits_{\ell \to \infty} \frac{k_{\ell}}{b_{\ell}} = b \in (0, \infty)$.
Let $A_{\ell}$ be a sequence that satisfies $\liminf_{\ell \to \infty} A_{\ell} = A$. Define $h_{\ell} (\cdot)$ on $[0, a_{\ell}]$ as
\begin{align}
h_{\ell} (w) = A_{\ell} \log (1 + Bw) - \frac{a_{\ell}}{k_{\ell}} H_2 \left(\frac{w}{a_{\ell}}\right).
\end{align}
Let $w_{\ell}^{\ast}$ achieve the global minimum of $h_{\ell} (\cdot)$ restricted to $[\bar{w}, b_{\ell}]$. For large enough $\ell$, either $ w_{\ell}^{\ast} = \bar{w}$ or $w_{\ell}^{\ast} \in [ c b_{\ell}, b_{\ell}]$, where
\begin{align}
c = \min \left\{\frac{b A}{64 (1 + A a)},1 \right\}.
\end{align}
\end{lemma}
\begin{IEEEproof}
The function $h_{\ell} (w)$ is equal to the difference of two concave functions. Its first two derivatives on $(0,a_{\ell})$ are:
\begin{align}
\label{eqn:h_first_derivative}
h'_{\ell}(w) &=  \frac{A_{\ell} B}{1+Bw} + \frac{1}{k_{\ell}} \log \frac{w}{a_{\ell} -w}
\end{align}
and
\begin{align}
h''_{\ell}(w) &= \frac{a_{\ell}}{k_{\ell} w (a_{\ell} - w)} - \frac{A_{\ell} B^2}{  (1+Bw)^2}
\\
&= \frac{a_{\ell} g_{\ell}(w) }{ k_{\ell} w (a_{\ell} - w)(1+Bw)^2},
\end{align}
where
\begin{align}
g_{\ell}(w) &= (B^2 + k_{\ell} A_{\ell} B^2/a_{\ell})w^2 + (2 B - k_{\ell} A_{\ell} B^2)w + 1.
\end{align}


Due to~\eqref{eq:assump_identify}, $k_{\ell} \to \infty$ as $\ell \to \infty$. For large enough ${\ell}$, $g_{\ell} (0) = 1$, $g_{\ell}(1) = - A_{\ell} B^2 k_{\ell} + A_{\ell} B^2 k_{\ell} /a_{\ell} + (B+1)^2 <0$, and $g_{\ell}(a_{\ell}) = (B a_{\ell} +1)^2 >0 $. Moreover, the minimum of the quadratic function $g_{\ell} (w)$ is achieved at:
\begin{align}
v_{\ell} = \frac{k_{\ell} A_{\ell} B - 2}{2B(1 + k_{\ell} A_{\ell} /a_{\ell})}\,.
\end{align}
Since $\frac{1}{2} k_{\ell} A_{\ell} B {\geq_{\ell}} 2$, we have $ k_{\ell} A_{\ell} B - 2 {\geq_{\ell}} \frac{1}{2} k_{\ell} A_{\ell} B$. Also, $A_{\ell} k_{\ell} / a_{\ell} \leq_{\ell} 1 + 2 A a$. We have
\begin{align}
\frac{v_{\ell} }{b_{\ell} } & {\geq_{\ell}} \frac{\frac{1}{2} \frac{k_{\ell}}{b_{\ell}} A_{\ell} B }{2 B (1 + A_{\ell} \frac{k_{\ell}}{a_{\ell}} )} \\
& {\geq_{\ell}} \frac{\frac{1}{2} \left(\frac{1}{2} b \right) \left( \frac{1}{2} A \right)}{2 (2+ 2 A a)} \\
\label{eq:vl}& = \frac{b A}{32( 1 + A a)} .
\end{align}

Note that $b_{\ell} \to \infty$ and~\eqref{eq:vl} implies $v_{\ell} \to \infty$. For large enough $\ell$, since $h''_{\ell} (w) < 0$ for every $w \in [\bar{w}, v_{\ell}]$, $h_{\ell} (w)$ is concave over $[\bar{w}, v_{\ell}]$. Since $v_{\ell} / b_{\ell} {\geq_{\ell}} 2 c$, we have either $w^*_{\ell} = \bar{w}$ or $w^*_{\ell} \in [c b_{\ell}, b_{\ell}]$ for large enough $\ell$.
\end{IEEEproof}

The general idea for proving Lemma~\ref{lemma:minhlambdarho} is to divide $\mathcal{W}^{(\ell)}$ into two regions based on whether the error probabily is dominated by false alarms or miss detections, and to lower bound $h_{\lambda,\rho}(w_1, w_2)$ given by~\eqref{eq:hlambdarho} for $(w_1, w_2)$ in those two regions separately. It is crucial to note that Lemma~\ref{lemma:minhlambdarho} claims the existence of a \textit{uniform} lower bound of $h_{\lambda,\rho}(w_1, w_2)$, i.e., $\ell^{\ast}$ is such that for every $\ell \geq \ell^{\ast}$, $h_{\lambda,\rho}(w_1, w_2) \geq c_0$ \textit{regardless} of $(w_1, w_2)$, which in general depend on $\ell$.
Define
\begin{align}\label{eq:theta_bar}
\phi_{\ell} = \frac{n (\ell)}{ k_{\ell} } = \frac{2 \ell H_2(\alpha_{\ell}) }{k_{\ell} \log(1+ k_{\ell} P') } ,
\end{align}
which can be regarded as the identification cost per active user. Let
\begin{align}
  \phi = \lim_{\ell \to \infty} \phi_{\ell},
\end{align}
which may be $\infty$. As $\phi \geq 0$, we prove the cases of $\phi >0$ and $\phi = 0$ separately.

\subsection{The case of $\phi > 0$}
\label{append:phineq0}

In this case, by~\eqref{eq:ident_n0}, the signature length is $n_0 = \left(1 + \epsilon \right) \phi_{\ell} k_{\ell}$. As we shall see, if the number of false alarms $w_2= |A \backslash A^{\ast}|$ is small, the error probability is dominated by miss detections; whereas for relatively large $w_2$, the error probability is dominated by false alarms.

Define the following positive constant:
\begin{align}\label{eq:bar_w}
\bar{w} = \max\left\{\frac{4}{P'} e^{(8 + 4 \epsilon) / \phi}, 1\right\} .
\end{align}
We will derive lower bounds of $h_{\lambda,\rho}(w_1, w_2)$ for the cases of $0 \leq w_2 \leq \bar{w}$ and $\bar{w} < w_2 \leq (1+\delta_{\ell}) k_{\ell}$ separately.

\subsubsection{The case of $0 \leq w_2 \leq \bar{w}$}

Recall that $\rho \in [0,1]$ and $\lambda \in [0,\infty)$ can be chosen arbitrarily to yield a lower bound. We shall always choose them to satisfy $0 \leq \lambda \rho \leq 1$. This implies that
\begin{align}
  \begin{split}
    &2 \log \left( 1 + \lambda(1 - \lambda \rho) w_2 P' + \lambda \rho (1 - \lambda \rho) w_1 P' \right) \geq \\
    & 
    \log \left( 1 + \lambda (1 - \lambda \rho) w_2 P' \right)   +
    \log \left( 1 +  \lambda \rho (1 - \lambda \rho) w_1 P' \right).
  \end{split}
\end{align}
In this case, a lower bound of $h_{\lambda,\rho}(w_1, w_2)$ can be splitted into two parts as
\begin{align}
\label{eq:lbhW1a} h_{\lambda,\rho}(w_1, w_2) & \geq g_{\lambda, \rho}^1 (w_1) + g_{\lambda, \rho}^2 (w_2),
\end{align}
where
\begin{align} \label{eq:g1}
  g_{\lambda,\rho}^1\! (w_1)
  = \frac{ n_0}{4 k_{\ell}} \log \left( 1 + \lambda \rho (1 - \lambda \rho) w_1 P' \right) -  \frac{|A^{\ast}|}{k_{\ell}} H_2 \Big( \!\frac{w_1}{ |A^{\ast}| } \!\Big)
\end{align}
and
\begin{align}  \label{eq:g2}
  \begin{split}
    & g_{\lambda, \rho}^2 (w_2) = \frac{n_0}{4 k_{\ell}} \log\left( 1 + \lambda (1 - \lambda \rho) w_2 P' \right) \\
    & \qquad - \frac{(1 - \rho ) n_0}{2 k_{\ell}} \log \left( 1 + \lambda w_2 P' \right) - \frac{\rho \ell}{k_{\ell}} H_2 \left( \frac{w_2}{\ell} \right).
  \end{split}
\end{align}

It is easy to see that $g_{\lambda, \rho}^1 (0) = g_{\lambda, \rho}^2 (0) = 0$. However, since $(w_1, w_2) \in \mathcal{W}^{(\ell)}$, $w_1$ and $w_2$ cannot be 0 simultaneously. In the following, we lower bound $g_{\lambda, \rho}^1 (w_1)$ for $w_1 \geq 1$ and $g_{\lambda, \rho}^2 (w_2)$ for $w_2 \geq 1$. Then $h_{\lambda,\rho}(w_1, w_2)$ can be lower bounded by the minimum of the two lower bounds of $g_{\lambda, \rho}^1 (w_1)$ and $g_{\lambda, \rho}^2 (w_2)$.

Choose  $\lambda =2/3$ and $\rho = 3/4$. We have
\begin{align} \label{eq:g_lb_casea}
  \begin{split}
    & g_{2/3, 3/4}^2 (w_2) = \frac{n_0}{4 k_{\ell}} \log\left( 1 + \frac{ w_2 P' }{3} \right) \\
    &\qquad - \frac{ n_0}{8 k_{\ell}} \log \left( 1 + \frac{2 w_2 P'}{3} \right) - \frac{3 \ell} {4 k_{\ell}} H_2 \left(\frac{w_2}{\ell} \right) .
  \end{split}
\end{align}
Since $(1+x)^{r} \leq 1 +r x$ for $r\in [0, 1]$, we have
\begin{align}
\label{eq:logconcave} \log(1 +r x ) \geq r \log(1+ x)
\end{align}
for $x \geq 0$ and the equality is achieved only if $x=0$. Letting $r = 1/2$, $x = 2 w_2 P'/3$, we can see that for $w_2 >0$,
\begin{align}
\log \left( 1 + \frac{ w_2 P'}{3} \right) > \frac{1}{2} \log \left( 1+ \frac{2 w_2 P'}{3} \right).
\end{align}
Define a positive constant
\begin{align}
\epsilon' = \min_{1 \leq w_2 \leq \bar{w}} \frac{\phi}{8} \left[\log \left( 1 + \frac{ w_2 P'}{3} \right) - \frac{1}{2} \log \left( 1+ \frac{2 w_2 P'}{3} \right) \right].
\end{align}
By Lemma~\ref{lemma:lim_H2}, $\frac{\ell} {k_{\ell}} H_2 (\bar{w}/ \ell)$ vanishes as $\ell$ increases. We can find some $\ell_0 > 2 \bar{w}$ such that for every $\ell \geq \ell_0$, $\phi_{\ell} > \phi/2$ and $\frac{3 \ell} {4 k_{\ell}} H_2 (\bar{w}/ \ell) \leq \frac{\epsilon'}{2}$.
For every $\ell \geq \ell_0$, we have $H_2(w_2/\ell) \leq H_2(\bar{w} / \ell)$ for $1 \leq w_2 \leq \bar{w}$ and thus $g_{2/3, 3/4}^2 (w_2)$ is lower bounded as
\begin{align}
\nonumber & g_{2/3, 3/4}^2 (w_2) \\
\nonumber &\geq \frac{\phi_{\ell}}{4 }  \left[\log \left( 1 + \frac{ w_2 P'}{3} \right) - \frac{1}{2} \log \left( 1+ \frac{2 w_2 P'}{3} \right) \right] \\
&\qquad - \frac{3 \ell} {4 k_{\ell}} H_2 (\bar{w}/ \ell) \\
&\geq \epsilon' - \frac{\epsilon'}{2} \\
\label{eq:lb_gw2_limgtr0}& = \frac{\epsilon'}{2}.
\end{align}

Meanwhile,
\begin{align}  \label{eq:lb_g_1}
  \begin{split}
    & g_{2/3, 3/4}^1 (w_1) = \\
    &\quad \frac{ (1+\epsilon) \phi_{\ell} }{4} \log \left( 1 + \frac{ w_1 P'}{4} \right) -  \frac{|A^{\ast}|}{k_{\ell}} H_2 \left( \frac{w_1}{|A^{\ast}|}  \right).
  \end{split}
\end{align}
When $w_1 \geq 1$, we shall invoke Lemma~\ref{lemma:lbhlargeK} to show that the minimum of the RHS of~\eqref{eq:lb_g_1} is achieved at either $w_1 = 1$ or some value close to $k_{\ell}$. Define
\begin{align}\label{eq:a0}
a = \min \left\{ \frac{ \phi}{16} \log\left( 1 + \frac{P'}{4} \right) , 1\right\}.
\end{align}

We consider the following three cases separately:
\begin{enumerate}[~~~~{Case} a)]
\item $1 \leq |A^{\ast}| \leq a k_{\ell}, 1 \leq w_1 \leq |A^{\ast}|$, 
\item $a k_{\ell} \leq |A^{\ast}| \leq (1+\delta_{\ell}) k_{\ell},  a k_{\ell}/2 \leq w_1 \leq |A^{\ast}|$, 
\item $a k_{\ell} \leq |A^{\ast}| \leq (1+\delta_{\ell}) k_{\ell},  1 \leq w_1 \leq a k_{\ell}/2$.
\end{enumerate}

For every $\ell \geq \ell_0$, $g_{2/3, 3/4}^1 (w_1)$ in Case (a) is lower bounded as
\begin{align}
 g_{2/3, 3/4}^1 (w_1) &\geq \frac{ \phi_{\ell} }{4} \log \left( 1 + \frac{ P'}{4} \right) - a \\
 & \geq  \frac{  \phi }{8} \log \left( 1 + \frac{ P'}{4} \right) - a \\
\label{eq:gw1casea} & \geq  \frac{  \phi }{16} \log \left( 1 + \frac{ P'}{4} \right).
\end{align}

In Case (b), $g_{2/3, 3/4}^1 (w_1)$ is lower bounded as
\begin{align}\label{eq:gw1caseb}
g_{2/3, 3/4}^1 (w_1) &\geq \frac{ (1+\epsilon) \phi_{\ell} }{4} \log \left( 1 + \frac{ a k_{\ell} P'}{8} \right) - (1+\delta_{\ell}),
\end{align}
which grows without bound as $\ell$ increases.

In Case (c), $w_1 / |A^{\ast}| \leq 1/2$. Since $H_2(\cdot)$ is increasing on $[0,1/2]$, by~\eqref{eq:lb_g_1},
\begin{align}
\nonumber & g_{2/3, 3/4}^1 (w_1) \\
&\geq \frac{ (1+\epsilon) \phi_{\ell} }{4} \log \left( 1 + \frac{ w_1 P'}{4} \right) -  \frac{(1+\delta_{\ell}) k_{\ell}}{k_{\ell}} H_2 \left( \frac{w_1}{a k_{\ell}}  \right) \\
\label{eq:lb_g_1a}& \geq \frac{2}{a} \left[ \frac{ (1+\epsilon)a \phi_{\ell} }{8} \log \left( 1 + \frac{ w_1 P'}{4} \right) -   \frac{a k_{\ell}}{k_{\ell}} H_2 \left( \frac{w_1}{a k_{\ell}}  \right) \right].
\end{align}
Applying Lemma~\ref{lemma:lbhlargeK} with $A_{\ell} = (1+\epsilon) a \phi_{\ell} /8$, $B = P'/4$,  $a_{\ell} = a k_{\ell}$, $\bar{w} =1$ and $b_{\ell} = a k_{\ell}/2$, we conclude that there exists $\ell_1$ such that for every $\ell \geq \ell_1$, the RHS of~\eqref{eq:lb_g_1a} restricted to $w_1 \in [1,a k_{\ell}/2]$ achieves the minimum either at $1$ or on $ [c a k_{\ell}/2,a k_{\ell}/2]$ for some $c\in (0,1]$. Moreover, $H_2 \left( \frac{1}{a k_{\ell}}  \right)$ vanishes as $\ell$ increases. There exists some $\ell_2$ such that for every $\ell \geq \ell_2$, $H_2 \left( \frac{1}{a k_{\ell}}  \right) \leq  \frac{\phi}{32} \log \left( 1 + \frac{P'}{4} \right)$ and $\phi_{\ell} \geq \phi/2$.

For every $\ell \geq \max \{\ell_1, \ell_2\}$, if the minimum of the RHS of~\eqref{eq:lb_g_1a} is achieved at $1$, then $g_{2/3, 3/4}^1 (w_1)$ in Case (c) is lower bounded as
\begin{align}
g_{2/3, 3/4}^1 (w_1) &\geq  \frac{\phi_{\ell}}{4} \log \left( 1 + \frac{P'}{4} \right)  -  2 H_2 \left( \frac{1}{a k_{\ell}}  \right) \\
&\geq  \frac{\phi}{8} \log \left( 1 + \frac{P'}{4} \right)  -  2 H_2 \left( \frac{1}{a k_{\ell}}  \right) \\
\label{eq:lb_g_w1}&\geq \frac{\phi}{16} \log \left( 1 + \frac{P'}{4} \right) .
\end{align}

For every $\ell \geq \max \{\ell_1, \ell_2\}$, if the minimum of the RHS of~\eqref{eq:lb_g_1a} is achieved on $[c a k_{\ell}/2,a k_{\ell}/2]$, then then $g_{2/3, 3/4}^1 (w_1)$ in Case (c) is lower bounded as
\begin{align}\label{eq:lb_g_w1_2}
g_{2/3, 3/4}^1 (w_1) \geq \frac{\phi_{\ell}}{4} \log \left( 1 + \frac{c a k_{\ell} P'}{8} \right) - 2,
\end{align}
which grows without bound as $\ell$ increases.

By~\eqref{eq:gw1casea},~\eqref{eq:gw1caseb},~\eqref{eq:lb_g_w1} and~\eqref{eq:lb_g_w1_2}, it concludes that for every $\ell \geq \max \{\ell_0, \ell_1, \ell_2 \}$,
\begin{align}
  g_{2/3, 3/4}^1 (w_1) \geq \frac{\phi}{16} \log \left( 1 + \frac{P'}{4} \right)
\end{align}
for every $1 \leq w_1 \leq |A^{\ast}|$ and for every $1 \leq |A^{\ast}| \leq (1+\delta_{\ell}) k_{\ell}$. Combining the lower bound of $g_{2/3, 3/4}^1 (w_2)$ given by~\eqref{eq:lb_gw2_limgtr0}, we conclude that for every $\ell \geq \max(\ell_0, \ell_1, \ell_2)$ and for every $(w_1, w_2) \in \mathcal{W}^{(\ell)}$ with $0 \leq w_2 \leq \bar{w}$, $h_{2/3,3/4}(w_1, w_2)$ can be uniformly lower bounded as
\begin{align}\label{eq:h_lb_case_1}
h_{2/3,3/4}(w_1, w_2) \geq & \min \left\{ \frac{\epsilon'}{2}, \frac{  \phi }{16} \log \left( 1 + \frac{ P'}{4} \right) \right\} .
\end{align}

\subsubsection{The case of $\bar{w} < w_2 \leq (1+\delta_{\ell}) k_{\ell} $}

Letting $\lambda = 1/2$ and $\rho = 1$ in~\eqref{eq:hlambdarho}, and using the fact that $w_1 \geq 0$ and $|A^{\ast}| / k_{\ell} \leq 2$, we have
\begin{align}
  \nonumber & h_{1/2,1}(w_1, w_2) \\
  & \geq \frac{ n_0 }{2 k_{\ell}} \log \bigg( 1 + \frac{w_2P'}{4} \bigg) - \frac{ \ell}{k_{\ell}} H_2 \Big( \frac{w_2}{\ell} \Big)
-  \frac{ |A^{\ast}|}{k_{\ell}} H_2  \Big( \frac{w_1}{|A^{\ast}|} \Big)  \\
\label{eq:lbhW2b} & \geq \frac{(1+\epsilon) \phi_{\ell} }{2 } \log \left( 1 + \frac{w_2 P'}{4} \right)  - \frac{\ell}{k_{\ell}} H_2 \left( \frac{w_2}{\ell} \right) - 2.
\end{align}

Applying Lemma~\ref{lemma:lbhlargeK} with $A_{\ell} = (1+\epsilon) \phi_{\ell} /2$, $B = P'/4$, $a_{\ell} = \ell$ and $b_{\ell} = (1 + \delta_{\ell}) k_{\ell}$, we can conclude that there exists some $\ell_3$ such that for every $\ell \geq \ell_3$, the minimum of the RHS of~\eqref{eq:lbhW2b} restricted to $[\bar{w},(1+ \delta_{\ell}) k_{\ell}]$ is achieved either at $\bar{w}$ or on $ [c k_{\ell}, (1 + \delta_{\ell}) k_{\ell}]$, for some $c \in (0,1]$. Moreover, by Lemma~\ref{lemma:lim_H2}, there exists some $\ell_4$ such that for every $\ell \geq \ell_4$, $\frac{ \ell}{k_{\ell}} H_2 (\bar{w}/ \ell) \leq 1$ and $\phi_{\ell} > \phi/2$.

For every $\ell \geq \max\{\ell_3, \ell_4\}$, if the minimum of the RHS of~\eqref{eq:lbhW2b} is achived at $\bar{w}$, then $h_{1/2, 1 }(w_1, w_2)$ is uniformly lower bounded as
\begin{align}
\label{eq:lbhW2bStep2_0} h_{1/2, 1 }(w_1, w_2) & \geq \frac{\phi }{4 } \log \left( 1 + \frac{\bar{w} P'}{4} \right)  - 2 \\
\label{eq:h12_0} & \geq \epsilon.
\end{align}

For every $\ell \geq \max\{\ell_3, \ell_4\}$, if the minimum of the RHS of~\eqref{eq:lbhW2b} is achieved on $[c k_{\ell}, (1 + \delta_{\ell}) k_{\ell}]$, we consider two cases:
\begin{enumerate}[~~~~{Case} a)]
\item $\ell > 2(1+\delta_{\ell}) k_{\ell}$, 
\item $\ell \leq 2(1+\delta_{\ell}) k_{\ell}$.
\end{enumerate}

In Case (a), $w_2 / \ell < 1/2$. Since $H_2(\cdot)$ is increasing on $[0,1/2]$, by~\eqref{eq:lbhW2b}, we have
\begin{align}
\nonumber & h_{1/2,1}(w_1, w_2) \\
&\geq \frac{(1+\epsilon) \phi_{\ell} }{2 } \log \left( 1 + \frac{c  k_{\ell} P'}{4} \right)  - \frac{ \ell}{k_{\ell}} H_2 \left(  \frac{(1+\delta_{\ell}) k_{\ell}}{\ell} \right) - 2 \\
\label{eq:lbhW2bStep1_0}& \geq \frac{(1+\epsilon) \phi_{\ell} }{2 } \log \left( 1 + \frac{c  k_{\ell} P'}{4} \right)  - (1 + \delta_{\ell}) \frac{ \ell}{k_{\ell}} H_2 \left(  \frac{ k_{\ell}}{\ell} \right) - 2 \\
\nonumber & = \frac{ \phi_{\ell} }{2} \left[ (1 + \epsilon) \log \left( 1 + \frac{ c k_{\ell} P'}{4} \right)  - (1 +  \delta_{\ell}) \log(1 +k_{\ell} P') \right] \\
\label{eq:lbhW2bStep1}&\hspace{7cm} - 2,
\end{align}
where~\eqref{eq:lbhW2bStep1_0} follows from~\eqref{eq:H2_ineq}, and~\eqref{eq:lbhW2bStep1} is due to~\eqref{eq:theta_bar}. By~\eqref{eq:deltal}, $\delta_{\ell} \log(1 +k_{\ell} P')$ vanishes as $k_{\ell}$ increases. Moreover,
\begin{align}
\lim_{k_{\ell} \to \infty} \log \left( 1 + \frac{ c k_{\ell} P'}{4} \right)  -  \log(1 +k_{\ell} P') = \log\frac{c}4.
\end{align}
Thus, the RHS of~\eqref{eq:lbhW2bStep1} grows without bound (uniformly for $(w_1, w_2)$) as ${\ell}$ increases.

In Case (b), by~\eqref{eq:lbhW2b}, we have
\begin{align}
h_{1/2,1}(w_1, w_2) & \geq \frac{(1+\epsilon) \phi_{\ell} }{2 } \log \left( 1 + \frac{c  k_{\ell} P'}{4} \right)  - \frac{\ell}{k_{\ell}}  - 2 \\
\label{eq:lbhW2bStep2} &\geq \frac{(1 + \epsilon) \phi_{\ell} }{2 } \log \left( 1 + \frac{c k_{\ell} P'}{4} \right)  - 5 ,
\end{align}
which grows without bound (uniformly for $(w_1, w_2)$) as ${\ell}$ increases.

By~\eqref{eq:h12_0},~\eqref{eq:lbhW2bStep1} and~\eqref{eq:lbhW2bStep2}, we conclude that for every $\ell \geq \max \{\ell_3, \ell_4\}$,
\begin{align}\label{eq:h_lb_case_2}
h_{1/2,1}(w_1, w_2) \geq \epsilon
\end{align}
uniformly for all $0 \leq w_1 \leq |A^{\ast}|$, $\bar{w} \leq w_2 \leq (1+\delta_{\ell}) k_{\ell}$, and $1 \leq |A^{\ast}| \leq (1+\delta_{\ell}) k_{\ell}$.

Combining~\eqref{eq:h_lb_case_1} and~\eqref{eq:h_lb_case_2}, we conclude that Lemma~\ref{lemma:minhlambdarho} holds for the case of $\phi > 0$ with $\ell^{\ast} = \max\{\ell_0, \ell_1, \ell_2, \ell_3, \ell_4\} $.

\subsection{The case of $\phi = 0$}
\label{append:phieq0}

In this case, $n_0 = \epsilon k_{\ell}$ by~\eqref{eq:ident_n0}. We let $\lambda = 3/5$, $\rho = 5/6$. Note that~\eqref{eq:lbhW1a}--\eqref{eq:g2} remain true in this case.

Consider first $g_{3/5, 5/6}^2 (w_2)$. By~\eqref{eq:logconcave}, we have
\begin{align}
\log \left( 1 + \frac{3 w_2 P'}{10} \right)  \geq \frac{1}{2} \log \left( 1 + \frac{3 w_2 P'}{5} \right) .
\end{align}
Thus,
\begin{align}
  \nonumber
  g_{3/5, 5/6}^2 (w_2)
  &= \frac{\epsilon}{4}
    \log \left(\! 1 + \frac{3 w_2 P'}{10} \!\right)
    - \frac{\epsilon}{12} \log \left(\! 1 + \frac{3 w_2 P'}{5} \!\right) \nonumber\\
  & \qquad\qquad- \frac{ 5 \ell}{6 k_{\ell}} H_2 \left( \frac{w_2}{\ell} \right)   \\
  \geq&\, \frac{\epsilon}{24 } \log \left( 1 + \frac{3 w_2 P'}{5} \right) - \frac{ 5 \ell}{6 k_{\ell}} H_2 \left( \frac{w_2}{\ell} \right).
  \label{eq:g2_lb}
\end{align}
Applying Lemma~\ref{lemma:lbhlargeK} with $A_{\ell} = \epsilon/20$, $B = 3 P'/5$, $\bar{w}=1$, $a_{\ell} = \ell$ and $b_{\ell} = (1+\delta_{\ell}) k_{\ell}$, we conclude that there exists some $\ell_5$ such that for every $\ell \geq \ell_5$, the minimum of the RHS of~\eqref{eq:g2_lb} restricted to $w_2 \in [1, (1+\delta_{\ell}) k_{\ell}]$ is achieved at either $1$ or on $[c k_{\ell}, (1 +  \delta_{\ell}) k_{\ell} ] $ for some $c \in (0,1]$. Moreover, by Lemma~\ref{lemma:lim_H2}, there exists some $\ell_6$ such that for every $\ell \geq \ell_6$, $\frac{ 5 \ell}{6 k_{\ell}} H_2 \left( \frac{1}{\ell} \right) \leq \frac{\epsilon}{48 } \log \left( 1 + \frac{3 P'}{5} \right)$.

For every $\ell \geq \max \{\ell_5, \ell_6\}$, if the minimum of the RHS of~\eqref{eq:g2_lb} is achieved at $1$, then $ g_{3/5, 5/6}^2 (w_2)$ is lower bounded as
\begin{align}
 g_{3/5, 5/6}^2 (w_2) &\geq \frac{\epsilon}{24 } \log \left( 1 + \frac{3 P'}{5} \right) - \frac{ 5 \ell}{6 k_{\ell}} H_2 \left( \frac{1}{\ell} \right) \\
\label{eq:gw2_case0_0} &\geq \frac{\epsilon}{48 } \log \left( 1 + \frac{3 P'}{5} \right).
\end{align}

For every $\ell \geq \max \{\ell_5, \ell_6\}$, if the minimum of the RHS of~\eqref{eq:g2_lb} is achieved on $[c k_{\ell}, (1 +  \delta_{\ell}) k_{\ell} ]$, we consider two cases:
\begin{enumerate}[~~~~{Case} a)]
\item $\ell > 2 (1+ \delta_{\ell}) k_{\ell}$, 
\item $\ell \leq 2 (1+ \delta_{\ell}) k_{\ell}$.
\end{enumerate}

In Case (a), $w_2 / \ell <1/2$. Since $H_2(\cdot)$ is increasing on $[0,1/2]$, we have
\begin{align}
\nonumber & g_{3/5, 5/6}^2 (w_2)   \\
& \geq \frac{\epsilon}{24 } \log \left( 1 + \frac{3 c k_{\ell} P'}{5} \right) -  \frac{ 5 \ell}{6 k_{\ell}} H_2 \left( \frac{(1+\delta_{\ell}) k_{\ell}}{ \ell} \right) \\
\label{eq:g2w2_lb1_0} & \geq \frac{\epsilon}{24 } \log \left( 1 + \frac{3 c k_{\ell} P'}{5} \right) - (1+\delta_{\ell}) \frac{ 5 \ell}{6 k_{\ell}} H_2 \left( \frac{ k_{\ell}}{ \ell} \right) \\
 & = \frac{\epsilon}{24 } \log \left(\! 1 + \frac{3 c k_{\ell} P'}{5} \! \right)
 \!-\! (1+\delta_{\ell}) \frac{5 \phi_{\ell} }{12} \log \left( 1+ k_{\ell} P \right) \\
 \label{eq:g2w2_lb1}
 & =\left[ \frac{\epsilon}{24 } - (1+\delta_{\ell}) \frac{5 \phi_{\ell} }{12}  \frac{\log \left( 1+ k_{\ell} P \right)}{\log \left( 1 + \frac{3 c k_{\ell} P'}{5} \right) } \right]  \log \left(\! 1 + \frac{3 c k_{\ell} P'}{5} \!\right)
\end{align}
where~\eqref{eq:g2w2_lb1_0} is due to~\eqref{eq:H2_ineq}.
Since $\phi_{\ell} \to 0$, we have
\begin{align}
 (1+\delta_{\ell}) \frac{5 \phi_{\ell} }{12}  \frac{\log \left( 1+ k_{\ell} P \right)}{\log \left( 1 + \frac{3 c k_{\ell} P'}{5} \right) } \to 0.
\end{align}
The RHS of~\eqref{eq:g2w2_lb1} thus grows without bound (uniformly for all $w_2$) as $\ell$ increases.

In Case (b), we have
\begin{align}
g_{3/5, 5/6}^2 (w_2)  & \geq \frac{\epsilon}{24 } \log \left( 1 + \frac{3 c k_{\ell} P'}{5} \right) - \frac{ 5 \ell}{6 k_{\ell}}  \\
\label{eq:g2w2_lb2} & \geq  \frac{\epsilon}{24 } \log \left( 1 + \frac{3 c k_{\ell} P'}{5} \right) -  \frac{10}{3} .
\end{align}
which grows without bound (uniformly for all $w_2$) as $k_{\ell}$ increases.

By~\eqref{eq:gw2_case0_0},~\eqref{eq:g2w2_lb1}, and~\eqref{eq:g2w2_lb2}, we conclude that for every $\ell \geq \max \{\ell_5, \ell_6\}$,
\begin{align}\label{eq:lb_gw2_lim0}
g_{3/5, 5/6}^2 (w_2) \geq \frac{\epsilon}{48 } \log \left( 1 + \frac{3 P'}{5} \right)
\end{align}
holds uniformly for all $1 \leq w_2 \leq (1+\delta_{\ell}) k_{\ell}$.

Consider next $g_{3/5, 5/6}^1 (w_1)$.
\begin{align}
g_{3/5, 5/6}^1 (w_1) = \frac{\epsilon}{4} \log \left( 1+ \frac{w_1 P'}{4} \right) - \frac{|A^{\ast}|}{k_{\ell}} H_2 \left( \frac{w_1}{|A^{\ast}|} \right).
\end{align}
Define
\begin{align}
a = \min \left\{\frac{\epsilon}{8} \log \left( 1 + \frac{P'}{4} \right), 1\right\} .
\end{align}

We consider the following three cases:
\begin{enumerate}[~~~~{Case} a)]
\item $1 \leq |A^{\ast}| \leq a k_{\ell}, 1 \leq w_1 \leq |A^{\ast}|$, 
\item $a k_{\ell} \leq |A^{\ast}| \leq (1+\delta_{\ell}) k_{\ell}, a k_{\ell} / 2 \leq w_1 \leq |A^{\ast}|$, 
\item $a k_{\ell} \leq |A^{\ast}| \leq (1+\delta_{\ell}) k_{\ell}, 1 \leq w_1 \leq a k_{\ell}/2$.
\end{enumerate}

In Case (a), $g_{3/5, 5/6}^1 (w_1)$ is uniformly lower bounded as
\begin{align}
g_{3/5, 5/6}^1 (w_1) &\geq \frac{\epsilon}{4} \log \left( 1+ \frac{ P'}{4} \right) - a \\
\label{eq:gw1_case0_0} & \geq \frac{\epsilon}{8} \log \left( 1+ \frac{ P'}{4} \right).
\end{align}

In Case (b),  $g_{3/5, 5/6}^1 (w_1)$ is uniformly lower bounded as
\begin{align}
\label{eq:gw1_case0_1} g_{3/5, 5/6}^1 (w_1) \geq \frac{\epsilon}{4} \log \left( 1+ \frac{a k_{\ell}  P'}{8} \right) - (1+\delta_{\ell}),
\end{align}
which grows without bound as $k_{\ell}$ increases.

In Case (c), $w_1 / |A^{\ast}| \leq 1/2$. Since $H_2(\cdot)$ is increasing on $[0,1/2]$, we have
\begin{align}
  g_{3/5, 5/6}^1 (w_1)
  &\geq  \frac{\epsilon}{4} \log \left( 1+ \frac{w_1 P'}{4} \right) - (1+\delta_{\ell}) H_2 \left( \frac{w_1}{a k_{\ell}} \right) \\
\label{eq:gw1_case0}
  \geq&\, \frac{\epsilon}{4} \log \left( 1+ \frac{w_1 P'}{4} \right) - \frac{2}{a} \frac{a k_{\ell}}{k_{\ell}} H_2 \left( \frac{w_1}{a k_{\ell}} \right) .
\end{align}
Applying Lemma~\ref{lemma:lbhlargeK} with $A_{\ell} =a \epsilon/8$, $B = P'/4$, $a_{\ell} = a k_{\ell}$, $\bar{w} = 1$ and $b_{\ell} = a k_{\ell}/2$, we conclude that there exists some $\ell_7$ such that for every $\ell \geq \ell_7$, the RHS of~\eqref{eq:gw1_case0} restricted to $w_1 \in [1, a k_{\ell} / 2]$ achieves minimum either at $1$ or on $[c a k_{\ell}/2, a k_{\ell}/2]$ for some $c \in (0,1]$. Moreover, there exists some $\ell_8$ such that for every $\ell \geq \ell_8$, $H_2 \left( \frac{1}{a k_{\ell}} \right) \leq \frac{\epsilon}{16} \log \left( 1+ \frac{ P'}{4} \right)  $.

For every $\ell \geq \max \{\ell_7, \ell_8\}$, if the minimum of the RHS of~\eqref{eq:gw1_case0} is achieved at $w_1 = 1$, then $g_{3/5, 5/6}^1 (w_1)$ in Case (c) is lower bounded as
\begin{align}
 g_{3/5, 5/6}^1 (w_1) &\geq \frac{\epsilon}{4} \log \left( 1+ \frac{ P'}{4} \right) - 2 H_2 \left( \frac{1}{a k_{\ell}} \right) \\
\label{eq:gw1_case0_2} &\geq \frac{\epsilon}{8} \log \left( 1+ \frac{ P'}{4} \right) .
\end{align}

For every $\ell \geq \max \{\ell_7, \ell_8\}$, if the minimum is achieved on  $[c a k_{\ell}/2, a k_{\ell}/2]$, then $g_{3/5, 5/6}^1 (w_1)$ in Case (c) is uniformly lower bounded as
\begin{align}
\label{eq:gw1_case0_3} g_{3/5, 5/6}^1 (w_1) \geq \frac{\epsilon}{4} \log \left( 1+ \frac{ a c k_{\ell} P'}{8} \right) - 2,
\end{align}
which grows without bound as $k_{\ell}$ increases.

By~\eqref{eq:gw1_case0_0},~\eqref{eq:gw1_case0_1},~\eqref{eq:gw1_case0_2} and~\eqref{eq:gw1_case0_3}, it concludes that for every $\ell \geq \max \{\ell_7, \ell_8\}$,
\begin{align}
g_{3/5, 5/6}^1 (w_1) \geq \frac{\epsilon}{8} \log \left( 1+ \frac{ P'}{4} \right)
\end{align}
holds uniformly for all $1 \leq w_1 \leq |A^{\ast}|$. Combining the lower bound of $g_{3/5, 5/6}^2 (w_2)$ given by~\eqref{eq:lb_gw2_lim0}, we conclude that for every $\ell \geq \max \{\ell_5, \ell_6, \ell_7, \ell_8\}$,  and every $1 \leq |A^{\ast}| \leq (1+\delta_{\ell}) k_{\ell}$,
\begin{align}
  \begin{split}
    &h_{2/3,3/4} (w_1, w_2) \\
    &\geq \min \left\{  \frac{\epsilon}{48} \log \left( 1 + \frac{3 P'}{5} \right) , \right. \left. \frac{\epsilon}{8} \log \left( 1 + \frac{P'}{4} \right) \right\}
  \end{split}
\end{align}
holds uniformly for all $(w_1, w_2) \in \mathcal{W}^{(\ell)}$.
Consequently, Lemma~\ref{lemma:minhlambdarho} is established for the case of $\phi = 0$. Combining the results of Appendix~\ref{append:phineq0} and Appendix~\ref{append:phieq0} proves Lemma~\ref{lemma:minhlambdarho}.

\section{Proof of Lemma~\ref{lemma:errexp} }
\label{append:pflemmaerr}

The lemma was proved for $k_n = o(n)$ in~\cite{chen2013gaussian}. In this paper, we prove the achievability result for $k_n = O(n)$. Throughout the proof, we focus on the case where $k_n$ grows without bound as $n$ increases, because the case of bounded $k_n$ was included in~\cite{chen2013gaussian}.

Let $f(\gamma,\rho)$ be defined as~\eqref{eq:f_gamma_rho}.
Choosing $\rho=1$, we have
\begin{align}
\nonumber & f(\gamma,1) = \\
& \frac{1}{2} \log\left( 1 + \frac{\gamma k_n P'}{2} \right) - \frac{(1-\epsilon) \gamma}{2} \log(1+k_n P') - \frac{k_n}{n} H_2(\gamma).
\end{align}
Denote $c_n = k_n /n$ and $c = \limsup_{n \to \infty} c_n$. By differentiating $f(\gamma,1)$ with respect to $\gamma$, we have
\begin{align}
  \begin{split}
    & \frac{d f(\gamma, 1)}{d \gamma} = \\
    &\frac{k_n P'}{4 + 2 \gamma k_n P'} - \frac{1-\epsilon }{2} \log(1+ k_n P') + \frac{k_n}{n} \log \frac{\gamma}{1 - \gamma},
  \end{split}
\end{align}
and
\begin{align}
\frac{d^2 f(\gamma, 1)}{d \gamma^2} &= \frac{c_n}{\gamma (1 - \gamma)} - \frac{(k_n P')^2}{2 (2+ \gamma k_n P')^2}.
\end{align}

Note that $k_n = O(n)$, $k_n$ is increasing without bound and $\gamma \geq 1/k_n$. Evidently,
\begin{align}
8 c_n &\leq_n k_n P'^2 /4 \\
&\leq \frac{1}{4} (k_n P')^2 \gamma.
\end{align}
Therefore, for sufficiently large $n$,
\begin{align}
8 c_n k_n P' \gamma  + 8 c_n \leq \frac{1}{2} (k_n P')^2 \gamma
\end{align}
 holds \textit{uniformly} for all $\gamma \in [1 /k_n, 1]$. Thus, for sufficiently large $n$,
\begin{align}
\nonumber & \frac{d^2 f(\gamma, 1)}{d \gamma^2} \\
&=  \frac{ (1+2 c_n) \gamma^2 (k_n P')^2 - (k_n P')^2 \gamma + 8 c_n k_n P' \gamma  + 8 c_n    }{2 (2+ \gamma k_n P')^2\gamma (1 - \gamma)} \\
&\leq \frac{ (1+2 c_n) \gamma^2 (k_n P')^2 - (k_n P')^2 \gamma + \frac{1}{2} (k_n P')^2 \gamma    }{2 (2+ \gamma k_n P')^2\gamma (1 - \gamma)} \\
& = \frac{ \left[(1+2 c_n) \gamma - 1/2 \right] (k_n P')^2 }{2 (2+ \gamma k_n P')^2  (1 - \gamma)} \\
\label{eq:f2order}& \leq \frac{ \left[(1+ 4c) \gamma - 1/2 \right] (k_n P')^2  }{2 (2+ \gamma k_n P')^2  (1 - \gamma)}
\end{align}
holds uniformly for all $\gamma$.

We pick the constant $\gamma' =  \frac{1/2}{1+ 4 c}$. Since $0 \leq c < \infty$, we have $0< \gamma' \leq 1/2$. By~\eqref{eq:f2order}, for sufficiently large $n$, $\frac{d^2 f(\gamma, 1)}{d \gamma^2} <0$ holds uniformly for all $1/ k_n \leq \gamma \leq \gamma'$. It means $f(\gamma,1)$ is concave over $\gamma \in [1/k_n, \gamma']$. Therefore, there exists some $N_0$ such that for every $n \geq N_0$,
\begin{align}\label{eq:min_f}
  \min_{1/k_n \leq \gamma \leq 1} f(\gamma,1)
  = \min \left\{f\left(\frac1{k_n},1\right), \min_{\gamma' \leq \gamma \leq 1} f(\gamma,1) \right\} .
\end{align}

If the minimum is achieved at $\gamma = 1/k_n$, we have
\begin{align}
  \begin{split}
    f\left(\frac1{k_n},1\right)
    &= \frac{1}{2} \log\left( 1 + \frac{ P'}{2} \right)
      - \frac{k_n}{n} H_2 \left( \frac{1}{k_n} \right) \\
    &\qquad - \frac{1-\epsilon}{2 k_n} \log(1+k_n P') .
  \end{split}
\end{align}
Since $(1/ k_n)\log(1+k_n P')$ and $\frac{k_n}{n} H_2(1/ k_n)$ vanishes as $k_n$ increases, there exists $N_1$ such that for every $n \geq N_1$,
\begin{align}\label{eq:lb_f_0}
f(1/k_n,1) \geq \frac{1}{4} \log\left( 1 + \frac{ P'}{2} \right).
\end{align}

If the minimum is achieved on $[\gamma',1]$, we have 
\begin{align}
  \begin{split}
    & f(\gamma,1) \geq \\
\label{eq:lb_f_1}& \frac{1}{2} \log\left( 1 + \frac{\gamma' k_n P'}{2} \right) - \frac{1-\epsilon}{2} \log(1+k_n P') - \frac{k_n}{n}.
  \end{split}
\end{align}
Since $ \log\left( 1 + \gamma' k_n P'/2 \right) - \log(1+k_n P')$ and $k_n / n$ converge to some constants, the lower bound given by~\eqref{eq:lb_f_1} grows without bound as $n$ increases.

In summary, combining~\eqref{eq:min_f},~\eqref{eq:lb_f_0}, and~\eqref{eq:lb_f_1}, it concludes that for every $n \geq \max\{N_0, N_1\}$ and every $|A^{\ast}|$, the error exponent is lower bounded
\begin{align}
E_r &\geq \min_{1/k_n \leq \gamma \leq 1} f(\gamma,1) \\
&\geq \frac{1}{4} \log\left( 1 + \frac{ P'}{2} \right).
\end{align}
The lemma is thus established.

\section{Proof of Theorem~\ref{thm:achiev_bounded_k}}
\label{sec:proof_bounded_k}

Unlike the case of unbounded $k_n$, there is a nonvanishing probability that the number of active users is zero.
Let $A^{\ast}$ denote the set of active users and $\Ecal_d$ denote the event of detection error. Given an increasing sequence $s_n$ satisfying the conditions specified in Theorem~\ref{thm:achiev_bounded_k}. The overall error probability can be calculated as
\begin{align} \label{eq:overall_err}
  \begin{split}
    \prob \left\{\Ecal_d \right\} \leq \prob
    & \left\{ |A^{\ast}| >  s_n \right\} +  \prob \left\{ \Ecal_d | 1 \leq |A^{\ast}| \leq  s_n \right\} \\
    &+ \prob \left\{ \Ecal_d |\, |A^{\ast}| =0  \right\}.
  \end{split}
\end{align}

By the Chernoff bound for binomial distribution~\cite{arratia1989tutorial}, the probability that the number of active users is greater than $s_n$ is calculated as
\begin{align}
\prob \left\{ |A^{\ast}|   > s_n \right\}  \leq  \exp \left( - k_n (s_n / k_n -1)^2 /3 \right) ,
\end{align}
which vanishes as $s_n$ grows without bound.

Note that the sequence $s_n$ satisfies $\ell_n e^{- \delta s_n} \to 0$ for all $\delta >0$ and
\begin{align}\label{eq:cond_sn}
\lim_{n \to \infty} \frac{ 2 s_n H_2 (s_n/\ell_n)  }{n \log(1 + s_n  P) } <1,
\end{align}
which are the regularity conditions for unbounded $k_n$ as specified in Case 1 of Theorem~\ref{thm:capacityKMac}. The error probability $\prob \left\{ \Ecal_d | 1 \leq |A^{\ast}| \leq  s_n \right\}$ vanishes by following exactly the same as the analysis for the case of unbounded $k_n$ (i.e., Case 1) by treating $s_n$ as an unbounded $k_n$.

It remains to analyze the identification error conditioned on $|A^{\ast}| = 0$. If no user is active, the received signal in the first $n_0$ channel uses is purely noise, i.e., $\bY^a = \bZ^a$. By the user identification rule~\eqref{eq:decode} with $k_n$ replaced by $s_n$, a detection error occurs if at least one user is claimed to be active. The conditional detection error probability can be calculated as
\begin{align} \label{eq:lb_err_supp0}
  \begin{split}
    \prob
    & \left\{ \Ecal_d |\, |A^{\ast}| =0  \right\} \leq \\
    &\sum_{w=1}^{(1+\delta_n) s_{n}} \binom{\ell_n}{w} \prob \left\{ \bigg\| \bZ^a - \sum_{i = 1}^w \bS_i^a \bigg\|^2 \leq \|\bZ^a\|^2  \right\}.
  \end{split}
\end{align}
Let $\bar{\bS} = \sum_{i = 1}^w \bS_i^a$. The entries of $\bar{\bS}$ are i.i.d. according to $\mathcal{N} (0, w P')$. We have
\begin{align}
  \nonumber \prob
  & \left\{ \bigg\| \bZ^a - \sum_{i = 1}^w \bS_i^a \bigg\|^2 \leq \|\bZ^a\|^2  \right\} \\
  &\quad= \prob \left\{ \sum_{i=1}^{n_0} Z_i^a \bar{S}_i \geq  \frac{1}{2} \| \bar{\bS} \|^2  \right\} \\
\label{eq:energy_gauss}
  &\quad= \Exp \left\{ \prob \left\{ \sum_{i=1}^{n_0} Z_i^a \bar{S}_i \geq  \frac{1}{2} \| \bar{\bS} \|^2  \right\} \bigg| \bar{\bS} \right\} .
\end{align}
Conditioned on $\bar{\bS}$, $\sum_{i=1}^{n_0} Z_i^a \bar{S}_i \sim \mathcal{N}(0, \|\bar{\bS}\|^2)$. Therefore,
\begin{align}
  \Exp \left\{ \prob \left\{ \sum_{i=1}^{n_0} Z_i^a \bar{S}_i \geq
  \frac{\| \bar{\bS} \|^2}2  \right\} \bigg| \bar{\bS} \right\}
  &\leq \Exp\left\{ \mathsf{Q} \left( \frac{\|\bar{\bS}\|}{2} \right) \right\} \\
  \label{eq:ub_Q}
  &\leq \Exp\left\{ e^{-\frac{\|\bar{\bS}\|^2}{8}} \right\} \\
  \label{eq:chis_moment}
  &= \left(1 + \frac{w P'}4\right)^{- \frac{n_0}{2}}
\end{align}
where~\eqref{eq:ub_Q} is due to $\mathsf{Q} (x) 
\leq e^{- x^2/2}$, and~\eqref{eq:chis_moment} follows because $\|\bar{\bS} \|^2/(wP)$ is chi-squared distributed with $n_0$ degrees of freedom.
Combining~\eqref{eq:lb_err_supp0},~\eqref{eq:energy_gauss}, and~\eqref{eq:chis_moment}, the detection error probability for $|A^{\ast}| = 0$ can be upper bounded as
\begin{align}
  \begin{split}
    & \prob \left\{ \Ecal_d |\,  |A^{\ast}| =0  \right\} \\
    & \leq\!\! \sum_{w=1}^{ (1+\delta_n) s_n}
    \!\! \exp\left[ \ell_n H_2\left(\frac{w}{\ell_n}\right)
    - \frac{n_0}{2} \log \left(1 + \frac{w P'}4\right) \right].
  \end{split}
\end{align}

Let $\theta_n$ be given by~\eqref{eq:gammaT} with $k_n$ replaced by $s_n$ and define $\theta = \lim_{n \to \infty} \theta_n$. By the choice of the signature length given by~\eqref{eq:NT0}, $n_0 \geq_n \delta n$, where $\delta = \min(\epsilon, \theta (1+\epsilon)/2 )$. For large enough $n$, the error probability can be further upper bounded:
\begin{align}\label{eq:err_prob_ub_a0}
\prob \left\{ \Ecal_d |\, |A^{\ast}| =0  \right\} \leq \sum_{w=1}^{(1+\delta_n) s_n} \exp\left( - s_n h(w) \right),
\end{align}
where
\begin{align}
h(w) = \frac{\delta n}{2 s_n} \log (1 + w P'/4) - \frac{\ell_n}{s_n} H_2 \left( \frac{w}{\ell_n} \right).
\end{align}
Note that $s_n = O(n)$. Applying Lemma~\ref{lemma:lbhlargeK} with $\ell = n$, $\bar{w} = 1$, $A_n = \delta n /(2 s_n)$, $k_n = s_n$, $a_n = \ell_n$ and $b_n = (1+\delta_n) s_n$, we conclude that for large enougn $n$, the minimum of $h(w)$ restricted to $[1, (1+\delta_n) s_n]$ is achieved either at $ 1$ or $[c s_n, (1+\delta_n) s_n]$ for some $0 < c \leq 1$.
As long as $s_n$ satisfies the conditions as specified in Theorem~\ref{thm:achiev_bounded_k}, $(\ell_n/s_n) H_2 \left( 1/\ell_n \right)$ vanishes as $n$ increases by Lemma~\ref{lemma:lim_H2}. For large enough $n$, if the minimum of $h(w)$ is achieved at $w = 1$, $h(w)$ is uniformly lower bounded by some constant $c_0 >0$. If the minimum of  $h(w)$ is achieved on $[c s_n, (1+\delta_n) s_n]$, it implies that $h(w)$ grows without bound. It concludes that there exists some $N_0$, such that for every $n \geq N_0$, $h(w)$ is uniformly lower bounded by $c_0$ for all $1 \leq w \leq (1+\delta_n) s_n$.

By~\eqref{eq:err_prob_ub_a0}, there exists some $N_0$ and $c_0 >0$ such that for every $n \geq N_0$,
\begin{align}\label{eq:vanish_err_noactive}
\prob \left\{ \Ecal_d |\, |A^{\ast}| =0  \right\} \leq (1+\delta_n) s_n e^{- c_0 s_n}.
\end{align}
Therefore, $\prob \left\{ \Ecal_d |\, |A^{\ast}| =0  \right\}$ vanishes as the blocklength $n$ increases. Since the three terms in the RHS of~\eqref{eq:overall_err} all vanish, the overall detection error probability also vanishes.

\section{Proof of Lemma~\ref{lemma:err_succ_equal}}
\label{append:proof_err_succ_equal}

Since the users adopt Gaussian random codes, by treating the other users as interference, the first user to be decoded effectively sees Gaussian noise with variance $1+ (k_n-1)P$. In order to prove the lemma, we show that the error probability of \textit{any} $\left( \lceil \exp(v(n)) \rceil, n \right)$ code for the first user, where the message length $v(n)$ is given by~\eqref{eq:msg_len_succ}, is lower bounded by some positive constant.

Let $\prob_m(v(n), n)$ denote the average error probability for the first user achieved by the best channel code of blocklength $n$ with message length $v(n)$, where each codeword satisfies the \textit{maximal} power constraint~\eqref{eq:powerconst}. Let $\prob_e(v(n), n)$ denote the average error probability for the first user achieved by the best channel code of blocklength $n$ with message length $v(n)$, where each codeword satisfies the \textit{equal} power constraint, i.e., each codeword lies on a power-sphere $\sum_{i = 1}^n s_{ki} = n P$. According to~\cite[eq. (83)]{shannon1959probability}, we have
\begin{align}\label{eq:pe_ineq}
\prob_m(v(n-1), n-1) \geq \prob_e(v(n-1), n).
\end{align}
We will lower bound $\prob_e(v(n-1), n)$ in order to show that $\prob_m(v(n), n)$ is strictly bounded away from zero for $v(n)$ given by~\eqref{eq:msg_len_succ}.

Let $\lambda >1$ be an arbitrary constant. Following the notations in~\cite[eq. (13)]{polyanskiy2010channel_dispersion}, let the decoding threshold be $\gamma = (n-1)(1 - \lambda \epsilon) C$, $P'_{\bY}$ be the distribution of $n$ i.i.d. Gaussian random variables with zero mean and variance $1+k_n P$, $P_{\bY | \bX = [\sqrt{P}, \cdots, \sqrt{P}]}$ be the distribution of $n$ i.i.d. Gaussian random variables with mean $\sqrt{P}$ and variance $1 + (k_n-1)P$, and $\beta_{1 - \epsilon_n} \left( P_{\bY | \bX = [\sqrt{P}, \cdots, \sqrt{P}]}, P'_{\bY} \right)$, where $\beta_{\alpha} (P,P')$ is the minimum error probability of the binary hypothesis test under hypothesis $P'$ if the error probability under hypothesis $P$ is not larger than $1- \alpha$.
The error probability $\prob_e(v(n-1), n)$ is lower bounded as (see also~\cite[eq. (88)]{polyanskiy2010channel_dispersion})
\begin{align}
\nonumber & \prob_e(v(n-1), n) \geq  \prob \left\{ \frac{1}{2 (1+Q)} \sum_{i=1}^n Q  (1 - Z_i^2) + 2 \sqrt{Q} Z_i \right. \\
\label{eq:lb_error_succ}& \qquad \quad  \leq - \lambda \epsilon n C -  (1-\lambda \epsilon) C \Bigg\}   - e^{-  (\lambda -1) (n-1) \epsilon C}.
\end{align}

We will follow a similar step as in~\cite{polyanskiy2010channel_dispersion} to further calculate the RHS of~\eqref{eq:lb_error_succ}.
Let $X_i =  -Q  (1 - Z_i^2) - 2 \sqrt{Q} Z_i$, where $Z_i$ are i.i.d. standard Gaussian random variables. Then $E X_i = 0$. By Rozovsky's large deviation result~\cite[Theorem 5]{polyanskiy2010channel_dispersion}, we have
\begin{align}
\prob \left\{ \sum_{i = 1}^n X_i > x \sqrt{S} \right\} \geq \mathsf{Q} (x) e^{- \frac{ d_1 T x^3}{ S^{3/2} }} \left( 1 - \frac{ d_2 T x}{S^{3/2}} \right),
\end{align}
where $d_1$ and $d_2$ are some universal constants, $S = \sum_{i = 1}^n E |X_i|^2 $, and $T = \sum_{i = 1}^n E |X_i|^3 $ which is equivalent to~\eqref{eq:succ_T}.

Then the first term in~\eqref{eq:lb_error_succ} can be calculated as
\begin{align}
  \begin{split}
    & \prob \left\{ \frac{1}{2 (1+Q)} \sum_{i=1}^n Q  (1 - Z_i^2) + 2 \sqrt{Q} Z_i \leq \right. \\
    & \quad  - \lambda \epsilon n C - (1- \lambda \epsilon) C \Bigg\}
    = \prob \left\{ \sum_{i=1}^n X_i \geq x \sqrt{S} \right\},
  \end{split}
\end{align}
where $x = {2 (\lambda \epsilon n + 1 - \lambda \epsilon) C (1+Q)}/{ \sqrt{S}} $.


We can derive that $S = 2 n Q (2+Q)$. Since $Q = P/(1+(k_n-1) P) \to 0$ as $n$ increases, we have
\begin{align}
  \Exp \left\{ |X_i|^3 \right\} = O \left( Q^{3/2 } \right).
\end{align}
Moreover, since $k = a n$, we have $T = O \left( n Q^{3/2} \right) $ and therefore $T$ tends to zero as $n$ increases.

\end{document}